\documentclass[12pt,twoside,singlespace]{mitthesis}

\usepackage{amssymb}
\usepackage{amsfonts}
\usepackage{amsmath}
\usepackage{amsthm}
\usepackage{latexsym}

\usepackage{afterpage}

\newcommand{\ket}[1]{|#1  \rangle}
\newcommand{\bra}[1]{\langle#1 |}
\newcommand{\proj}[1]{|{#1} \rangle\langle {#1} |}
\newcommand{\braket}[2]{\langle #1 | #2 \rangle}
\newcommand{\set}[1]{\left\{ {#1} \right\}}
\newcommand{\paren}[1]{\left( {#1} \right)}
\newcommand{\floor}[1]{\left\lfloor{#1} \right\rfloor}
\newcommand{\ceil}[1]{\left\lceil{#1} \right\rceil}
\newcommand{\abs}[1]{\left| {#1} \right|}
\newcommand{\supp}{\mathrm{supp}}

\newcommand{\myspan}{\mathrm{span}}

\newcommand{\ith}{i^{th}}
\newcommand{\jth}{j^{th}}

\newcommand{\lth}{\ell^{th}}

\newcommand{\tr}{\mbox{Tr}}
\newcommand{\ky}{\ket{\psi}}
\newcommand{\kf}{\ket{\phi}}
\newcommand{\bray}{\bra{\psi}}
\newcommand{\braf}{\bra{\phi}}

\newcommand{\ze}{\mathbb{Z}}
\newcommand{\ce}{\mathbb{C}}

\newtheorem{theorem}{Theorem}[chapter]
\newtheorem{lemma}[theorem]{Lemma}
\newtheorem{prop}[theorem]{Proposition}

\newtheorem{fact}[theorem]{Fact}
\newtheorem{corollary}[theorem]{Corollary}

\newtheorem{definition}{Definition}

\theoremstyle{definition}

\newtheorem{protocol-num}{Protocol}
\newtheorem{algm}{Algorithm}
\newtheorem{rem}{Remark} 

\newenvironment{protocol}[1]{
      \begin{protocol-num}[#1]
        }
      {
      \end{protocol-num}
}

\newcommand{\lemref}[1]{Lemma~\ref{lem:#1}}
\newcommand{\factref}[1]{Fact~\ref{fact:#1}}
\newcommand{\propref}[1]{Proposition~\ref{prop:#1}}
\newcommand{\remref}[1]{Remark~\ref{rem:#1}}
\newcommand{\defref}[1]{Definition~\ref{def:#1}}
\newcommand{\chapref}[1]{Chapter~\ref{cha:#1}}
\newcommand{\secref}[1]{Section~\ref{sec:#1}}
\newcommand{\appref}[1]{Appendix~\ref{cha:#1}}
\newcommand{\stepref}[1]{Step~\ref{step:#1}}
\newcommand{\protref}[1]{Protocol~\ref{prot:#1}}
\newcommand{\algref}[1]{Algorithm~\ref{alg:#1}}
\newcommand{\figref}[1]{Figure~\ref{fig:#1}}

\newcommand{\tgood}{\mbox{2{\sc -good}}}
\newcommand{\vqss}{{\sc vqss}}
\newcommand{\css}{{\sc css}}
\newcommand{\cvss}{{\sc vss}}
\newcommand{\ftqc}{{\sc ftqc}}
\newcommand{\mpqc}{{\sc mpqc}}
\newcommand{\ttp}{{\cal TTP}}
\newcommand{\A}{{\cal A}}
\newcommand{\E}{{\cal E}}
\newcommand{\C}{{\cal C}}
\renewcommand{\H}{{\cal H}}
\newcommand{\TC}{{\tilde {{\cal C}}}}
\newcommand{\I}{{\cal I}}
\renewcommand{\O}{{\cal O}}
\newcommand{\Sim}{{\cal S}}

\newcommand{\ri}{\mathcal{R}^I}

\newcommand{\fn}{{\cal F}^{\otimes n}} 
\newcommand{\F}{{\cal F}}

\newcommand{\vq}{V^{(q)}} \newcommand{\vp}{V^\perp}
\newcommand{\vbp}{V_B^\perp} \newcommand{\vbq}{V_B^{(q)}}

\newcommand{\wq}{W^{(q)}} 
\renewcommand{\wp}{W^\perp}
\newcommand{\wbp}{W_B^\perp} \newcommand{\wbq}{W_B^{(q)}}

\newenvironment{nar-enum}{\noindent \begin{enumerate}
    \setlength{\leftmargin}{0pt} \setlength{\rightmargin}{0pt}
    \setlength{\parsep}{0pt} \setlength{\itemsep}{1pt}
    \setlength{\topsep}{0pt}} {\end{enumerate}}

\newenvironment{nar-item}{\noindent \begin{itemize}
    \setlength{\leftmargin}{0pt} \setlength{\rightmargin}{0pt}
    \setlength{\parsep}{0pt} \setlength{\itemsep}{1pt}
    \setlength{\topsep}{0pt}} {\end{itemize}}

\newenvironment{nar-desc}{\noindent \begin{description}
    \setlength{\leftmargin}{0pt} \setlength{\rightmargin}{0pt}
    \setlength{\parsep}{0pt} \setlength{\itemsep}{1pt}
    \setlength{\topsep}{0pt}} {\end{description}}

\renewenvironment{proof}{\noindent \textbf{Proof}:}{$\Box$ \vskip 12pt}
\newenvironment{proofof}[1]{\vskip 12pt \noindent \textbf{Proof} (of
  {#1}):}{$\Box$ \vskip 12pt}

\newcommand{\an}[1]{}

\begin{document}


%
%
%
%
%
%
%

\title{Multi-party Quantum Computation}

\author{Adam Smith}

\department{Department of Electrical Engineering and Computer Science}
\degree{Master of Science in Electrical Engineering and Computer Science}
\degreemonth{September}
\degreeyear{2001}
\thesisdate{August 22, 2001}

\prevdegrees{B.Sc. Mathematics and Computer Science\\
McGill University, 1999.}


\supervisor{Madhu Sudan}{Associate Professor}

\chairman{Arthur C. Smith}{Chairman, Department Committee on Graduate Students}

\maketitle



\cleardoublepage
\setcounter{savepage}{\thepage}
\begin{abstractpage}

%
%
%

We investigate definitions of and protocols for multi-party quantum
computing in the scenario where the secret data are quantum systems.
We work in the quantum information-theoretic model, where no
assumptions are made on the computational power of the adversary.  For
the slightly weaker task of \emph{verifiable quantum secret sharing},
we give a protocol which tolerates any $t < n/4$ cheating parties (out
of $n$).  This is shown to be optimal. We use this new tool to
establish that any multi-party quantum computation can be securely
performed as long as the number of dishonest players is less than
$n/6$.

This thesis is based on joint work with Claude Cr{\'e}peau and Daniel Gottesman.

\end{abstractpage}


\cleardoublepage


\pagestyle{plain}

\section*{Acknowledgements}

Thanks to my co-authors, Claude Cr\'epeau and Daniel Gottesman, for
their invaluable help with this research. Thanks to my eternally
patient supervisor, Madhu Sudan. Thanks also to the many students,
family mmbers, faculty and friends who provided encouragement, support
and advice.

\tableofcontents
\listoffigures

\chapter{Introduction}
\label{cha:intro}

  





    












Secure distributed protocols have been an important and fruitful area
of research for modern cryptography. In this setting, there is a group
of participants who wish to perform some joint task, despite the fact
that some of the participants in the protocol may cheat in order to
obtain additional information or corrupt the outcome. When we approach
distributed cryptography from the perspective of quantum computing,
a number of natural questions arise:
\begin{itemize}
\item Do existing classical protocols remain secure when the adversary
  has access to a quantum computer?
\item Can we use quantum computing and communication to find new, more
  secure or faster protocols for classical tasks?
\item What new, \emph{quantum} cryptographic tasks can we perform?
\end{itemize}

This research is inspired by the last of these questions. We propose
to investigate a quantum version of an extensively studied classical
problem, \emph{secure multi-party computation} (or \emph{secure
  function evaluation}), first introduced by \cite{GMW87}. In this
scenario, there are $n$ players in a network. Each player $i$ has an
input $x_i$, and the players want to run a protocol to collectively
compute some joint function $f(x_1,...,x_n)$. The challenge is that
all players would like this function evaluation to be \emph{secure}.
Informally, this means:
\begin{enumerate}
\item \emph{Soundness and Completeness:} At the end of the protocol,
  all honest players should learn the correct function value
  $f(x_1,...,x_n)$.
\item \emph{Privacy:} Cheating players should learn nothing at all
  beyond what they can deduce from the function output and their own
  inputs.
\end{enumerate}

\paragraph{Multi-party Quantum Computation}

For this thesis, we consider an extension of this task to quantum
computers. A multi-party quantum computing (\mpqc) protocol allows $n$
participants $P_{1},P_{2},\ldots,P_{n}$ to compute an $n$-input
quantum circuit in such a way that each party $P_{i}$ is responsible
for providing one (or more) of the input states. The output of the
circuit is broken in $n$ components ${\cal
  H}_{1}\otimes\ldots\otimes{\cal H}_{n}$ such that $P_{i}$ receives
the output ${\cal H}_{i}$.  Some components ${\cal H}_{i}$ may be
empty.

Note that the inputs to this protocol are arbitrary quantum
states---the player providing an input need only have it in his
possession, he does not need to know a classical description of
it%
\footnote{For quantum information, merely having a state in one's
  possession in not the same as knowing a description of it, since one
  cannot completely measure an unknown quantum state}. %
Moreover, unlike in the classical case, we cannot assume without loss
of generality that the result of the computation will be broadcast.
Instead, each player in the protocol receives some part of the output.

Informally, we require two security conditions as before. On one
hand, no coalition of $t$ or fewer cheaters should be able to affect
the outcome of the protocol beyond what influence they have by
choosing their inputs. On the other hand, no coalition of $t$ or fewer
cheaters should be able to learn anything beyond what they can deduce
from their initial knowledge of their input and from the systems ${\cal
  H}_i$ to which they have access. We formalize this notion in
\secref{def}.

\paragraph{Verifiable Quantum Secret Sharing} 

In order to construct \mpqc\ protocols, we consider a subtask which we
call \emph{verifiable quantum secret sharing}. In classical
cryptography, a verifiable secret sharing scheme \cite{ChorGMA85} is a
two-phase protocol with one player designated as the ``dealer''. After
the first phase (\emph{commitment}), the dealer shares a secret
amongst the players. In the second phase (\emph{recovery}), the
players reconstruct the value publicly. When the dealer passes the
first phase of the protocol, then
\begin{itemize}
\item \emph{Soundness:} There is a uniquely defined value $s$ which
  will be reconstructed in the second phase, regardless of any
  interventions by an adversary who can control no more than $t$
  players.
\item \emph{Completeness:} If the dealer is honest, then he always
  passes the commitment phase and the value $s$ recovered in the
  second phase is the secret he intended to share.
\item \emph{Privacy:} If the dealer is honest, no coalition of $t$
  players can learn any information about $s$.
\end{itemize}

The natural quantum version of this allows a dealer to share a state
$\rho$ (possibly unknown to him but nonetheless in his possession).
Because quantum information is not clone-able, we cannot require that the
state be reconstructed publicly; instead, the recovery phase also has
a designated player, the reconstructor $R$. We require that, despite
any malicious actions by a coalition of up to $t$ players:
\begin{itemize}
\item \emph{Soundness:} As long as $R$ is honest and the dealer passes
  the commitment phase successfully, then there is a unique quantum
  state which can be recovered by $R$.
\item \emph{Completeness:} When $D$ is honest, then he always passes
  the commitment phase. Moreover, when $R$ is also honest, then the
  value recovered by $R$ is exactly $D$'s input $\rho$.
\item \emph{Privacy:} When $D$ is honest, the adversaries learn
  no information about his input until the recovery phase.
\end{itemize}

Note that the privacy condition in this informal definition is
redundant, by the properties of quantum information: any information
adversaries could obtain about the shared state would imply some kind
of disturbance (in general) of the shared state, which would
contradict the completeness requirement. A formal definition of
security is given in \secref{def}.

\paragraph{Contributions}

The results of this thesis are based on unpublished joint work with
Claude Cr{\'e}peau and Daniel Gottesman \cite{CGS01}. In this thesis:
\begin{nar-item}
\item We give a protocol for verifiable quantum secret sharing that
  tolerates any number $t<n/4$ of cheaters.
\item We show that this is optimal, by proving that \vqss\ is
  impossible when $t\geq n/4$.
\item Based on techniques from fault-tolerant quantum computing, we
  use our \vqss\ protocol to construct a multi-party quantum
  computation protocol tolerating any $t<n/6$ cheaters.
\end{nar-item}

Our protocols run in time polynomial in both $n$, the number of
players, and $k$, the security parameter. The error of the protocols
(to be defined later) is exponentially small in $k$.

Beyond these specific results, there are a number of conceptual
contributions of this thesis to the theory of quantum cryptographic
protocols.

\begin{nar-item}
\item We provide a simple, general framework for defining and proving
  the security of distributed quantum protocols in terms of
  equivalence to an ideal protocol involving a third party. This
  follows the definitions for classical multi-party protocols, which
  have been the subject of considerable recent work
  \cite{GL90,Beaver91,MR91,Can00,DM00,CDDIM01,PW00,Can01,Graaf97}.
\item The analysis of our protocols leads us to consider various
  notions of local ``neighborhoods'' of quantum states, and more
  generally of quantum codes. We discuss three notions of a
  neighborhood. The notion most often used for the analysis of quantum
  error-correction and fault-tolerance is insufficient for our needs,
  but we show that a very natural generalization---specific to
  so-called ``\css'' codes, is adequate for our purposes.
\item Along the way, we provide modified versions of the classical
  sharing protocols of \cite{CCD88}. The key property these protocols
  have is that dealers do not need to remember the randomness they use
  when constructing shares to distribute to other players. This allows
  them to replace a random choice of coins with the
  \emph{superposition} over all such choices.
\end{nar-item}

\paragraph{Organization}

The thesis is organized as follows. \chapref{intro} contains the
material necessary for understanding the protocols of this thesis as
well as their context. \secref{prevwork} describes the previous work
on the topics in this thesis, with emphasis on the works whose results
we use directly.  In \secref{def}, we present a framework for defining
security of a distributed quantum protocol which involves interaction
with a trusted third party. We use this framework to formally define
both verifiable quantum secret sharing and multi-party quantum
computation.  \secref{prelim} contains the mathematical background for
understanding our protocols, as well as results we use from the existing
literature. In \secref{nbhds}, we introduce three definitions of the
local ``neighborhoods'' of a quantum code, in order to help the reader
understand exactly what properties our protocols guarantee and what
properties are needed in our security analyses. Some additional
relations between these three notions are shown in \appref{morenbhds}.

The protocols which are the main focus of this thesis are presented in
\chapref{protocols}. One of the main proof techniques we use is a
``quantum-to-classical reduction'' (terminology due to \cite{LC99kd}).
In \secref{SP}, we illustrate this technique with a simple protocol
which achieves \vqss\ for a small number of cheaters ($t<n/8$), and
whose analysis will prove insightful for the sequel.  \secref{2ls}
uses a similar technique, but applied to a modified version of the
classical ``verifiable blob'' protocol of \cite{CCD88}, to construct a
\vqss\ protocol secure against $t<n/4$ cheaters. In \secref{imposs},
we show this is optimal by relating \vqss\ protocols to
error-correcting codes and applying the quantum Singleton bound.
Finally, we use our sharing scheme to contruct \mpqc\ protocol which
tolerates any $t<n/6$ cheaters.

We conclude with some open questions related to our results
(\chapref{conc}).

\section{Previous Work}
\label{sec:prevwork}

\paragraph{Classical {\sc mpc}}

Most of the work on classical distributed protocols is based on
\emph{secret sharing}, in which a message is encoded and shared
amongst a group of players such that no coalition of $t$ players gets
any information at all about the encoded secret, but any group of
$t+1$ or more players can recover the secret exactly. The prototypical
and most commonly used solution to this is the polynomial sharing
scheme due to Shamir \cite{Shamir79}: choose a random polynomial $p$
of degree at most $t$ over $\ze_p$ (for some prime $p>n$) subject to
$p(0)=s$, where $s$ is the secret being shared. The share given to
player $i$ is value $p(i)$, for $i=1,...,n$. Note that for normal
secret sharing we assume that the shares are prepared honestly.

This assumption was removed in subsequent work: Multi-party computing,
in which no player may be assumed to be honest, was first treated
explicitly by Goldreich et al.  \cite{GMW87}, although the subtask of
verifiable secret sharing had been investigated previously by Chor et
al. \cite{ChorGMA85}.  Goldreich et al. \cite{GMW87} proved that
\emph{under computational assumptions}, secure multi-party evaluation
of any function was possible tolerating any minority of cheating
players, i.e. for any $t <\frac n 2$.

Subsequently, Ben-Or et al. \cite{BGW88} and Chaum et al.
\cite{CCD88} independently proved that tolerating up to $t<\frac n 3$
was possible \emph{without} computational assumptions, provided that
one assumed that every pair of participants was connected by a secure
channel. Moreover, this bound is tight due to the impossibility of
even agreeing on a single bit when $t\geq \frac n 3$ (see Lynch
\cite{Lynch}, for example). The main difference between the results of
\cite{CCD88} and those of \cite{BGW88} is that the former allow a small
probability of error (exponentially small in the complexity of the
protocol). 

The bound of $\frac n 3$ for information-theoretically secure {\sc
  mpc} was broken by Rabin and Ben-Or \cite{RB89} and Beaver
\cite{Beaver89}, who showed that assuming the existence of a secure
broadcast channel, then one can in fact tolerate any minority ($t<
\frac n 2$) of cheaters without computational assumptions. Their
protocols introduce a small error probability, which is provably
unavoidable \cite{RB89}. The results of \cite{RB89,Beaver89} were
extended to the model of adaptive adversaries by Cramer et al.
\cite{CDDHR99}.

All of these protocols rely on verifiable secret sharing. Our solution
draws most heavily on the techniques of \cite{CCD88}. The essential
idea behind their {\sc vss} protocol is to share the secret using a
two-level version of the basic scheme of Shamir (above), and then use
a cut-and-choose zero-knowledge proof to allow the dealer to convince
all players that the shares he distributed were consistent with a
single polynomial $p(x)$.

Beyond these basic protocols, a line of work has focused on coming up
with proper definitions of multi-party computing
\cite{GL90,Beaver91,MR91,Can00,DM00,CDDIM01,PW00,Can01}. Both
\cite{Can01} and \cite{CDDIM01} provide summaries of that literature.
Most of the research has focused on finding definitions which allow
composability of protocols, mainly focusing on multi-party computing
(often referred to, more precisely, as \emph{secure function
  evaluation}). In this work, we adopt a simple definition (based on
the initial definitions of Canetti).  We do not prove any composition
protocols, but simply ensure that the definition captures our
intuition of security and is provably achieved by our protocols.  See
\secref{def} for further discussion.

\paragraph{Multi-party Quantum Protocols}

Relatively little work exists on multi-party cryptographic protocols
for quantum computers.  Secret sharing with a quantum secret was first
studied by Cleve, Gottesman and Lo \cite{CGL99}. They suggested a
generalization of the Shamir scheme, which is also used by Aharonov
and Ben-Or \cite{AB99} as an error-correcting code. One of the
contributions of \cite{CGL99} was that to point out the strong
connection between secret sharing and error-correcting codes in the
quantum setting (see \secref{qss}). Our \vqss\ protocol is based on
the \cite{CGL99} scheme, using a modification of the techniques of
\cite{CCD88} to ensure the consistency of distributed shares.

There were some additional works on distributed quantum protocols.
Gottesman \cite{Gott00} showed that quantum states could be used to
share classical secrets more efficiently than is possible in a
classical scheme. Chau \cite{Chau00} proposed a scheme for speeding up
\emph{classical} multi-party computing using quantum techniques;
\cite{Chau00} also mentions the problem of verifiable quantum secret
sharing as an open question.  The dissertation of van de Graaf
\cite{Graaf97} discusses defining the security of classical
distributed protocols with respect to a quantum adversaries, but
contains no constructions.



\paragraph{Fault-tolerant Quantum Computing}

In our proposed solution, we also use techniques developed for
fault-tolerant quantum computing (\ftqc). The challenge of \ftqc\ is
to tolerate \emph{non-malicious} faults occurring within a single
computer. One assumes that at every stage in the computation, every
qubit has some probability $p$ of suffering a random error, i.e. of
becoming completely scrambled (this corresponds to the classical
notion of random bit flips occurring during a computation). Moreover,
errors are assumed to occur \emph{independently} of each other and of
the data in the computation.

One can view multi-party computation as fault-tolerant computing with a
different error model, one that is suited to distributed computing. On
one hand, the \mpqc\ model is weaker in some respects since we assume
that errors will always occur in the same, limited number of
positions, i.e. errors will only occur in the systems of the $t$
corrupted players.

On the other hand, the error model of \mpqc\ is stronger in some
respects: in our setting errors may be \emph{maliciously} coordinated.
In particular, they will not be independently placed, and they may in
fact depend on the data of the computation---the adversaries will use
any partial information known about the other players' data, as well
as information about their own data to attempt to corrupt the
computation. For example, several \ftqc\ algorithms rely on the fact
that at certain points in the computation, at most one error is likely
to occur. Such algorithms will fail in a model of adversarially placed
errors.

Techniques from \ftqc\ are nonetheless useful for multi-party
computing. Considerable research has been done on \ftqc. We rely
mainly on the techniques of Aharonov and Ben-Or \cite{AB99}, which
were based on those of Shor \cite{Shor96}. Using ``\css'' quantum
error-correcting codes, Shor showed that fault-tolerance was possible
so long as the error rate in the computer decreased logarithmically
with the size of the computation being performed. Aharonov and Ben-Or
showed that by using concatenated coding, one could in fact tolerate a
constant error rate. They also introduced generalized \css\ codes in
which the individual pieces of a codeword are assumed to be
higher-dimensional systems, such as collections of several qubits
(this corresponds to using larger alphabets in classical coding
theory). 


\paragraph{Provably Secure (and Insecure) Quantum Protocols} 

While quantum cryptographic protocols have existed for some time, many
of them have been proven secure only recently. The first proofs of
security appeared in the context of entanglement purification
protocols \cite{BBPSSW96,DEJMPS96,LC99kd}.  In a different line of
work, Mayers \cite{Mayers98} provided a notoriously difficult proof
that the Bennett-Brassard key distribution scheme was secure. Unifying
these two lines of research, Shor and Preskill \cite{SP00} proved the
correctness of the Bennett-Brassard \cite{BB84} key distribution
protocol, based on a previous proof of a purification-based protocol
due to Lo and Chau \cite{LC99kd}. The main insight of \cite{LC99kd}
was that in certain situations, proving the security of a quantum
protocol could be reduced to classical probability arguments, since
one could assume without loss of generality that the adversary
followed one of a finite number of classical cheating strategies (a
so-called ``quantum-to-classical reduction''). A similar technique is
used in \cite{BCGST01} to prove the correctness of a scheme for
authenticating quantum transmission.  This technique will also be
useful for proving the soundness of our protocol, as it will allow us
deal with possible entanglement between data and errors by
``reducing'' them to classical correlations.

Note that for protocols where the adversary is one of the participants
in the system and not an outside eavesdropper, much less is known.
Some proofs were also attempted for tasks such as bit commitment
\cite{BCJL93}, but those proofs were later discovered to be flawed,
since bit commitment was proven impossible
\cite{Mayers96,LC97spec,Mayers97,LC96gen,LC97,BCMS98}. There have also
been several works on quanutm coin-tossing. Although arbitrarily small
error is known to be impossible, several works have focused on
reducing the error as much as possible
\cite{LC96gen,MS99,ATVY00,Amb01}. Yet another line of work has focused
on how to achieve certain two-party tasks using computional
assumptions, i.e. assuming that there exist (quantum) one-way
permutations \cite{DMS00,CLS01}.

\clearpage
\section{Definitions}
\label{sec:def}

This section describes a simple framework for proving the security of
distributed quantum cryptographic protocols. The defintions are based
on the initial framework of Canetti \cite{Can00}, as well as on
discussions in the dissertation of van de Graaf \cite{Graaf97}. We
describe two models for protocols. The first one--the ``real''
model---describes the environment we ultimately expect our protocols
to run in. The second model is idealized model in which players can
interact with an incorruptable outside party.  We will prove our
``real-model'' protocols secure by showing that they are equivalent to
a simple protocol for the ideal model which captures our notion of
what security means for a given task.

We provide no general composition theorems in this work. Instead, we
simply prove the security of our composed protocols directly. 

\subsection{``Real'' Model for Protocols}

For the protocols in this paper, we assume that every pair of players
is connected by perfect (i.e. authenticated, secret) quantum and
classical channels. Moreover, we assume that there is a classical
authenticated broadcast channel to which all players have access.
Because we will consider settings where $t<\frac n 4 < \frac n 3$, we
can also assume that players can perform \emph{classical} multi-party
computations \cite{BGW88,CCD88}%
\footnote{In fact, even the assumption of a broadcast channel is
  unnecessary but (since $t<\frac n 3$) but is made for simplicity.}.

The adversary is an arbitrary quantum algorithm (or family of
circuits) $\mathcal{A}$. We make no assumptions about the
computational power of the adversary; he is limited only by the number
of players $t$ that he can corrupt.

The \emph{initial configuration} for the protocol is the joint
state $\rho$ of $n+2$ quantum systems: an input system
$\mathcal{I}_i$ for each player in the protocol ($i=1,...,n$), as
well as the adversary's auxiliary input system $\mathcal{I}_{aux}$
and an outside reference system $\mathcal{I}_{ref}$ (which will
remain untouched throughout the protocol). Note that the input can
be an arbitrary quantum state, possibly entangling all these
systems.

A run of a ``real model'' protocol begins with all players receiving
their input system ${\cal I}_i$ and the adversary receiving the
state ${\cal I}_{aux}$. The adversary then chooses a subset $C$ of
size at most $t$ of players to corrupt. From then on, the
adversary has access to the state of the players in $C$ and
controls what they send over the channels. The adversary may cause
the cheaters' systems to interact arbitrarily. His only restriction
is that he has no access to the state of the honest players, and
cannot intercept their communication. The reference system ${\cal
I}_{ref}$ is untouched during this process.

At the end of the protocol, all players produce an output (for honest
players, this is the output specified by the protocol). The system
output by player $i$ is denoted ${\cal O}_i$. Moreover, the adversary
outputs an additional system ${\cal O}_{aux}$. The \emph{output
  configuration} for the run of the protocol is the joint state of
$\O_1,...,\O_n$, the adversary's state $\O_{aux}$ and the reference
system $\I_{ref}$. This state depends on the adversary $\A$ and the
initial configuration $\rho$, and is denoted $Real(\A,\rho)$. Note
that this configuration does not include any ancillary states or
workspace used by honest players, only the output specified by the
protocol (i.e. all other parts of the honest players' systems are
``traced out'').

\subsection{``Ideal'' Model For Protocols}

The main difference of the ideal model from the real model is that
there is a trusted third party (denoted $\ttp$) who helps the players in
the execution of some protocol. The communications model is the same
as before, except that every player is connected to $\ttp$ via a
perfect (i.e. authentic, secret) quantum channel. There is no need to
assume a broadcast channel since players can simply give a classical
value to $\ttp$ and ask that it be re-sent to all players.

As before, the initial configuration consists of $n$ systems
$\I_i$ containing the players' inputs as well as the two systems
$\I_{aux}$ and $\I_{ref}$. The $\ttp$ gets no input. The protocol
proceeds as in the real model, except that players may interact
with the $\ttp$, who may not be corrupted by the adversary. Finally,
the output configuration is the same as before. The final state of
the $\ttp$ is not included in the output configuration. The output
configuration for adversary $\A$ and initial configuration $\rho$
is denoted $Ideal(\A,\rho)$.

\subsection{Protocol Equivalence}

Suppose we have a protocol $\pi$ which is supposed to implement
some ideal functionality $f$, that is $f$ is an ideal model
protocol and $\pi$ is an attempt to implement it in the real
model.

Informally, we say $\pi$ implements $f$ if the input/output
behavior of $\pi$ cannot be distinguished from that of $f$.
Formally:

\begin{definition}[Perfect security]
A protocol $\pi$ is considered perfectly secure if for all
adversaries $\A_1$, there exists an adversary $\A_2$, running in
time polynomial in that of $\A_1$, such that for all input
configurations $\rho$ (possibly mixed or entangled), we have:
$$Real(\A_1,\rho) = Ideal(\A_2,\rho)$$
\end{definition}

The protocols we design do not in fact achieve this strong notion
of security. Instead, they take a security parameter $k$ as input.
All players receive the classical string $1^k$ as part of their
input (in the ideal model, so does the $\ttp$). Moreover, the inputs
may additionally depend on $k$ (in particular, we allow the
adversary's auxiliary input to depend on $k$). Since honest
players should be polynomial-time quantum circuits, the protocol will
run in time polynomial in $k$, although the adversary need not.

\begin{definition}[Statistical security]\label{def:statsecur}
A protocol $\pi$ is considered statistically secure if for all
adversaries $\A_1$, there exists an adversary $\A_2$, running in
time polynomial in that of $\A_1$, such that for all
\emph{sequences} of input configurations $\set{\rho_k}$ (possibly
mixed or entangled), we have:
$$F\paren{Real(1^k,\A_1,\rho_k) ,\
Ideal(1^k,\A_2,\rho_k)} \geq 1- 2^{-k},$$
 where $F$ denotes the
fidelity of two quantum density matrices.
\end{definition}

\paragraph{Simulators}

Our definition asks us to construct a new adversary $\A_2$ for every
real adversary $A_1$. To do so, we will follow the standard
cryptographic paradigm of constructing a \emph{simulator} $\Sim$ who
uses $\A_1$ as a black box. Thus we can write $\A_2 = \Sim^{\A_1}$.\ 
We can view $\Sim$ as an ``interface'' between the real-world
adversary and the ideal-model protocol \cite{Graaf97}: $\Sim$
exchanges messages with $\A_1$, but must also control the corrupted
parties in the ideal-model protocol.

When $\A_2$ is constructed in this way, then the definition above can
be restated: Suppose that at the end of the protocol the adversary
gains access to the outputs of the honest players. There should not
exist a real-world adversary $\A_1$ that can tell the difference
between (a) a run of the real protocol and (b) a run of the
ideal-model protocol with $\Sim$ as an interface. We will construct
simulators for our protocols in \secref{l2-sim} and \secref{dc}.

\subsection{Static versus Adaptive Adversaries}
\label{sec:adapt-def}

In this thesis, we consider only \textrm{static} adversaries, who
choose the parties they will corrupt before the beginning of the
protocol and remain with that choice. On the other hand, an
\emph{adaptive} adversary chooses which players to corrupt as the
protocol is progressing. The set of corrupted parties is still
monotone---we do not allow a player to become honest again once he has
been corrupted%
\footnote{An adversary who corrupts players dynamically is called a
  mobile adversary, and protocols for handling such
  adversaries are called \emph{pro-active}.}%
---but the adversary can base his decision on the message he is seeing
in the protocol. For example, if the players were to elect a small
group of participants to make some decision amongst themselves, an
adaptive adversary could wait until the selection had been made and
then corrupt the members of that small group. Proving protocols secure
against adaptive adversaries has been problematic even in the
classical setting \cite{CFGN96,CDDHR99}.

Choosing to handle only static adversaries simplifies the definitions
and proofs considerably, and offers no real loss of intuition.
Nonetheless, we believe that the protocols we describe here are secure
against adaptive adversaries, assuming that the environment in which
the protocol is running somehow records which parties were corrupted
and in what order (it is unclear what adaptivity even means without
such an assumption). In \secref{sp-sound}, we discuss briefly how some
of the proofs could be extended to handle adaptivity (see
\remref{adapt-sound}, p.~\pageref{rem:adapt-sound}).

\subsection{Multi-party Quantum Computation}
\label{sec:mpc-def}

We define multi-party quantum computation by giving an ideal-model
protocol for that task. Simply put, all players hand their inputs to
the trusted party, who runs the desired circuit and hands back the
outputs. Note that the only kind of cheating which is possible is that
cheaters may choose their own input. In particular, cheaters cannot
force the abortion of the protocol. One possible extension of this
work is to consider protocols where cheaters may not compromise the
correctness of the computation but might force the protocol to stop
before completion (see Open Questions, \chapref{conc}).

\begin{figure}[h!]
  \begin{center}
    \fbox{\parbox[t]{\textwidth}{\small
        \begin{protocol}{Multi-party Quantum Computation---Ideal Model}
          \label{prot:mpqc-ideal}
          \begin{nar-desc}
            \item[Pre:] All players agree on a quantum circuit $U$
            with $n$ inputs and $n$ outputs(for simplicity, assume
            that the $\ith$ input and output correspond to player
            $i$).
          
            \item[Input:] Each player gets an input system $S_i$ (of
            known dimension, say $p$).
        \end{nar-desc}
        \begin{nar-enum}
        \item (\textbf{Input Sharing})  For each $i$, player $i$ sends
          $S_i$ to $\ttp$. If $\ttp$ does not receive anything, then he
          broadcasts ``Player $i$ is cheating'' to all players.
          Otherwise, $\ttp$ broadcasts ``Player $i$ is OK.''

          \item (\textbf{Computation}) $\ttp$ evaluates the
            circuit $U$ on the inputs $S_i$. For all $i$ who cheated,
            $\ttp$ creates $S_i$ in a known state (say $\ket 0$).
          \item (\textbf{Output})
            \begin{nar-enum}
            \item $\ttp$ sends $\ith$ output to player $i$. 
            \item Player $i$ outputs the system he receives from $\ttp$.
            \end{nar-enum}
          \end{nar-enum}
        \end{protocol}
        }}\caption{\protref{mpqc-ideal} (Multi-party Quantum
      Computation---Ideal Model)}
     \label{fig:mpqc-ideal}
  \end{center}
\end{figure}

\subsection{Verifiable Quantum Secret Sharing}
\label{sec:vqss-def}

Providing a definition verifiable quantum secret sharing is trickier
than it is for multi-party computing. The idea of the ideal protocol
is simple. In the sharing phase, the dealer gives his secret system to
the trusted party. In the reconstruction phase, the $\ttp$ sends the
secret system to the reconstructor $R$.

However, a problem arises because \vqss\ is a two phase task, and the
formalism we established in the preceding sections only describes
one-phase protocols, which have a simpler input/output behaviour. For
example, if all we required of \vqss\ is that the reconstructor's
output be the same as the dealer's input, we could simply have $D$
send his secret system to $R$ without violating the definition---a
clear indication that such a definition would be insufficient.  For
the purposes of this thesis, we adopt a simple modification of the
definition of the preceding sections which allows us to describe
\vqss: instead of giving all inputs to the parties at the beginning of
the run of the protocol, some inputs are not given to the parties
until the beginning of the reconstruction phase.

Specifically, two of the inputs are delayed. First, players learn the
identity of the reconstructor $R$ only at the beginning of the
reconstruction phase (note that this doesn't stop the adversary from
knowing $R$ since the definition requires security for all adversaries
and input sequences). Second, the adversary also receives a second
auxiliary input $\mathcal{I}^{(2)}_{aux}$ at the beginning of the
reconstruction.  This allows us to capture any side information gained
by the adversary during interactions which occur between the end of
the sharing phase and the beginning of the reconstruction phase.

The ideal-model protocol we obtain is given in \figref{vqss-ideal}.
The definition of security we will use for this two-phase model is
essentially the same as for the one-phase model. An input
configuration $\rho$ consists of player identities $D$ and $R$, a
secret system $S$ and the two auxiliary inputs $\mathcal{I}_{aux}$ and
$\mathcal{I}^{(2)}_{aux}$. We require that for all adversaries
$\A_1$, there exists an adversary $\A_2$ such that for all sequences
of input configurations $\set{\rho_k}_{k\in\mathbb{N}}$, the fidelity
of the output of the real protocol to the output of the ideal protocol
is exponentially close to 1.


\begin{figure}[h!] 
  \begin{center}
    \fbox{\parbox[t]{\textwidth}{\small
        \begin{protocol}{Verifiable Quantum Secret Sharing---Ideal Model}
          \label{prot:vqss-ideal}

          \begin{nar-item}
          \item \textbf{Sharing Phase:} 
            \begin{nar-enum}
            \item \textbf{Inputs:} All players get $D$'s identity.
              Dealer $D$ gets a qupit $S$ (i.e. a $p$-dimensional
              system, where $p$ is a publicly agreed-upon integer).\\
              (Adversary also gets his auxiliary input
              $\mathcal{I}_{aux}$.)
            \item $D$ sends the $p$-dimensional system $S$ to $\ttp$.
              If $D$ fails to send $S$, then $\ttp$ broadcasts ``$D$ is
              cheating'' to all players. Otherwise, $\ttp$ broadcasts
              ``OK''.
          \end{nar-enum}
        \item \textbf{Reconstruction Phase:} 
          \begin{nar-enum}
          \item \textbf{Inputs:} All players get $R$'s identity.\\
            (Adversary also gets his second auxiliary input
            $\mathcal{I}^{(2)}_{aux}$.)
          \item If $D$ did not cheat in the sharing phase, $\ttp$ sends
            $S$ to the receiver $R$.
          \end{nar-enum}
        \end{nar-item}
        \end{protocol}
        }} 

      \caption{\protref{vqss-ideal} (Verifiable Quantum Secret
        Sharing---Ideal Model)}
      \label{fig:vqss-ideal}
  \end{center}
\end{figure}

\clearpage
\section{Mathematical Preliminaries}
\label{sec:prelim}

We assume that the reader is familiar with the basic notation and
formalism of quantum computing. For an introduction, the reader should
refer to a textbook such as Nielsen and Chuang \cite{NC00}.

For most of this paper, we will work with ``qupits'', that is
$p$-dimensional quantum systems, for some prime $p$. It is natural to
view the elements of the field $F=\ze_p$ as a basis for the state space
of a qupit.

In our settings, it will be useful to choose $p$ so that $n<p$. We
need not choose $p$ very big for this, since there is always a prime
between $n$ and $2n$. However, all of our protocols will remain
polynomial time even when $p$ is exponential in $n$. That is, the
complexity of the protocols will be polynomial in $\log |F| = \log p$.

Just as for the case of qubits, there are a few natural operators on qupits
which we will use extensively in this paper.

The shift and phase operators for qupits (sometimes denoted
$\sigma_x,\sigma_z$) are defined analogously to the case of qubits:
$$X\ket a \mapsto \ket{a+1 \mod p} \quad \mbox{and} \quad Z\ket{a}
\mapsto \omega^{a}\ket{a},$$
where $\omega = e^{2\pi i /p}$. These two
operators generate the Pauli group. Since they have a simple
commutation relation ($XZ = \omega ZX$), any element of the group is
proportional to some product $X^xZ^z$ for $x,z \in \set{0,...,p-1}$.
As for qubits, the $p^2$ operators $X^xZ^z$ form a basis for the space
of $p \times p$ complex matrices, and so any unitary operator on
qupits can be written as a linear combination of Pauli matrices.  In
particular, this is useful since means that correcting Pauli errors in
a quantum code is sufficient for correcting arbitrary errors. In the
context of errors, $X$ is called a \emph{shift error} and $Z$ is a
\emph{phase error}.

For registers of qupits, the Pauli matrices are tensor products of
Pauli matrices acting on individual qupits. If ${\bf x}=(x_1,...,x_n)$
and ${\bf y}=(y_1,...,y_n)$ are vectors in $\ze_p^n$, then $X^{\bf
  x}Z^{\bf z}$ denotes $X^{x_1}Z^{z_1}\otimes \cdots \otimes
X^{x_n}Z^{z_n}$. These form a basis for the space of operators on the
register. The set of positions on which a Pauli matrix does \emph{not}
act as the identity is called its \emph{support}, and is equal to the
union of the supports of ${\bf x}$ and ${\bf z}$. The number of such
positions is called the \emph{weight} of the operator.

\paragraph{Fourier Rotations}

Another transformation which arises often is the Fourier transform on
qupits, which generalizes the Hadamard rotation on qubits.
$$\F \ket{a} \mapsto \sum_{b \in \ze_p} \omega^{ab}\ket{b}$$
This is
called a Fourier rotation since its effect on the $p$-dimensional
vector of coefficients of the state of a qupit is exactly that of the
Fourier transform over the group $\ze_p$. Consequently, phase changes
become shifts in this new basis, and conversely:
$$
\F X = Z \F \quad \mbox{and} \quad \F Z = X^{-1}\F$$

A useful property of the Fourier transform is that linear
transformations remain linear after the change of basis.
Specifically, let $V$ be an invertible $n\times n$ matrix over
$\ze_p$. Let $ V$ denote the corresponding unitary operator on a
register of $n$ qupits, i.e $\widetilde V\ket {\bf x} = \ket{V {\bf
    x}}$. Then in the Fourier basis, this looks like a different
linear map, given by the matrix $(V^{-1})^\top$. That is $\F
\widetilde V \F^{-1} = \widetilde {(V^{-1})^\top)}$.

The main feature we will use is simply that the transformation remains
a linear permutation of the basis vectors. There is one very useful
special case. For controlled addition (denoted $c\mbox{-}X$), which maps
$\ket{a,b}\mapsto\ket{a,a+b}$, conjugating by a Fourier rotation
yields another controlled-addition, applied in the opposite direction
and with a scaling factor of $-1$ (i.e.
$\ket{a,b}\mapsto\ket{a-b,b}$).

\subsection{Quantum Error-Correction}
\label{sec:qecc}

A quantum error-correcting code is a way of encoding redundancy into
quantum information to allow correction of errors which occur during
transmission or storage. An $[[n, k,d]]$ quantum code encodes $k$
qubits into $n$ qubits (for $n \geq k$) and corrects any (arbitrary)
error which affects less than $\frac d 2$ positions in the code.  The
most resilient quantum codes actually work over higher-dimensional
subspaces, i.e. each ``position'' in the code consists of a qupit.
Recall that we work with qupits of dimension $p$, where $p$ is some
prime greater than $n$.

\paragraph{{\sc Css} Codes}

An important family of quantum codes are the \css\ codes (due to
Calderbank-Shor \cite{CS96} and Steane \cite{Steane96}). A \css\ code
over $n$ qupits is defined by two classical linear codes $V$ and $W$
over $\ze_p$, both of length $n$. They are chosen such that $V^\perp
\subseteq W$, where $V^\perp$ is the dual of $V$ with respect to the
standard dot product $v\cdot w=\sum_{i=1}^n v_iw_i$. Note that we
automatically also have $W^\perp \subseteq V$.  The quantum code
${\cal C}$ is then the set of states $\ket \psi$ which would yield a
codeword of $V$ if they were measured in the computational basis
($\set{\ket 0, \ket 1,...,\ket {p-1}}$), and yield a codeword of $W$
if they were measured in the Fourier basis
($\set{\F\ket0,...,\F\ket{p-1}}$).
  
Now for any given system of $n$ qupits and any linear subspace $W\leq F^n$,
we define
$$W^{(q)} = \mathrm{span}\{\ket{{\bf w}}:\ {\bf w} \in W\}.$$ 
If we denote by $\F^{\otimes n}$ the parallel application of $\F$ to
all qubits of an $n$-qubit register, then we have:
$$\C = \vq \cap \F\wq$$

The dimension of $\C$ as a code, i.e. number of qupits it can encode,
is simply $\dim(V/W^\perp)=\dim V - \dim W^\perp$. For convenience, we
will denote $V_0=W^\perp$ and $W_0 = \vp$, and so the formula for the
number of qupits encoded becomes $\dim V - \dim V_0 = \dim W - \dim W_0$.

\paragraph{Minimum Distance}

To correct an arbitrary error on a subset $A$ of positions ($A
\subseteq \set{1,...,n}$), it turns out that it is sufficient (and
necessary) to be able to correct Pauli errors, i.e.  compositions of
shift and phase errors applied to the qupits in $A$. Thus, to correct
errors on any $t$ positions it suffices to correct all Pauli errors of
weight at most $t$. A sufficient condition is that the spaces
$\set{E\C}$ be mutually orthogonal, where $E$ ranges over all Pauli
operators of weight at most $t$. In such a case, one can correct any
of these errors $E$ on a codeword $\ket \psi$ by performing a
measurement that identifies which of these subspaces contains the
corrupted codeword $E \ket \psi$, and then applying the correction
$E^{-1}$. This can be rephrased: for all Pauli operators of weight
at most $2t$, $E\C$ and $\C$ should be orthogonal spaces.  The minimum
distance of a quantum code $C$ is thus the weight of the smallest
Pauli operator for which this is not true.

\begin{definition}
  The minimum distance of a quantum code $\C$ is the weight of the
  smallest Pauli operator such that $\C$ and $E\C$ are not orthogonal.
\end{definition}

By the previous discussion, a code with distance $d$ can correct
arbitrary errors on any $\floor{(d-1)/2}$ positions. For \css\ codes,
there is a simple way to calculate the minimum distance:

\begin{fact}
  Let $V,W$ be classical codes with minimum distances $d_1$ and $d_2$
  such that $V^\perp \subseteq W$. Then the quantum \css\ code $\C=\vq
  \cap \F\wq$ has minimum distance at least $\min(d_1,d_2)$.
  \footnote{In fact, the minimum distance of $\C$ is the minimum of
    the weights of the lightest vectors in $V-V_0$ and $W-W_0$. These
    are bounded below by the minimum distances $d_1,d_2$, and the
    bound is tight for the codes used in this paper.}
\end{fact}

\paragraph{Syndromes and Error Correction}

Given a classical linear code $V$ of dimension $k$, the syndrome for
$V$ is a linear function from $n$ bits to $n-k$ bits that indicates
which coset of $V$ contains its argument. If $V$ has distance at least
$2t+1$ and a codeword $\mathbf{v} \in V$ is altered in $t$ or fewer positions,
then the syndrome of the corrupted word $\mathbf{v+e}$ allows one to compute
the correction vector $\mathbf{-e}$. We will let $V$-syndrome denote the
syndrome with respect to $V$. Note that computing the $V$-syndrome is
easy. Fix a basis $\set{\mathbf{v}_1,...,\mathbf{v}_{n-k}}$ of the dual code $V^\perp$.
The $V$-syndrome of $\mathbf{w}$ is the vector $(\mathbf{v_1\cdot w},...,\mathbf{v}_{n-k}\cdot
\mathbf{w})$.

This is the basis for the error correction procedure for \css\
codes. Suppose that $E=X^{\mathbf x}Z^{\mathbf z}$, and both ${\bf x}$ and
${\bf z}$ have support on at most $t$ positions. Let
$\ket\psi\in\C$. Since $\ket \psi$ lies in $\vq$, measuring the
$V$-syndrome of $E\ket \psi$ (in the computational basis) allows one to
compute the vector ${\bf x}$, and apply the correction
$X^{\bf -x}$. Similarly, measuring the $W$-syndrome in the Fourier basis
allows one to compute ${\bf z}$ and apply $Z^{\bf -z}$, thus recovering
$\ket \psi$.  The two measurements commute, so in fact it does not
really matter which one is applied first. 

The pair of measurement results used, namely the $V$-syndrome in the
computational basis and the $W$-syndrome in the Fourier basis, are
referred to together as the \emph{quantum syndrome}. If the syndromes
are $s_1$ and $s_2$ pits long respectively, then there are
$p^{s_1+s_2}$ possible quantum syndromes. This divides the whole space
$\ce^{\ze_p^n}$ into $p^{s_1+s_2}$ orthogonal subspaces indexed by the
set of \emph{equivalence classes} of Pauli operators. That is, two
Pauli operators $E,E'$ are deemed equivalent if $E\C=E'\C$; and the
space $\ce^{\ze_p^n}$ can be written as the direct sum of the
orthogonal spaces $\set{E_j\C}_{j\in J}$, where $J$ is a set of
indices which contains exactly one element from each equivalence
class. For a \css\ code, two Pauli operators $X^{\bf x}Z^{\bf z}$ and
$X^{\bf x'}Z^{\bf z'}$ will be equivalent if and only if ${\bf x}$ and
${\bf x'}$ are in the same coset of $V$, and ${\bf z}$ and ${\bf z'}$
are in the same coset of $W$.

Note that the dimension of the code can also be written as $n-s_1-s_2$.

\paragraph{Quantum Reed-Solomon Codes}

In this work we will use a family of \css\ codes known as ``quantum
polynomial codes'' or ``quantum Reed-Solomon codes''. These were
introduced by Aharonov and Ben-Or \cite{AB99}, and generalize
classical Reed-Solomon codes. 

In this paper, we will specify a quantum RS code by a single parameter
$\delta< (n-1)/2$, which represents the degree of the polynomials used
in the code. The corresponding code $\mathcal{C}$ will encode a single
qupit and correct $t=\floor{\frac \delta 2}$ errors. 
For simplicity, choose $\delta =
2t$. We will always choose the number $n$ of players to be either
$2\delta+1$ or $3\delta +1$.

If $n$ is the number of players, choose any $p$ such that $p>n$ %
(\footnote{In fact, the construction can be changed to allow $p=n$.}).
We will work over the field $F=\ze_p$. The classical Reed-Solomon code
$V^{\delta}$ is obtained by taking the vectors
$$\mathbf{\hat{q}}=\left( q(1),q(2),\ldots,q(n)\right)$$
for all
univariate polynomials $q$ of degree at most $\delta$. The related
code $V^{\delta}_0$ is the subset of $V^\delta$ corresponding to
polynomials which interpolate to 0 at the point 0. That is:
\begin{eqnarray*}
  V^\delta &=&\{ \mathbf{\hat q}:\ q \in F[x] :\ \deg(q) \leq \delta\} \\
  V_0^\delta &=& \{ \mathbf{\hat q}:\ \deg(q) \leq \delta 
  \mbox{ and }q(0)=0\} \subseteq V^\delta
\end{eqnarray*}
The code $V^{\delta}$ has minimum distance $d=n-\delta$. Moreover,
errors (up to $\floor{(n-\delta-1)/2}$ of them) can be corrected
\emph{efficiently}, given the syndrome of the corrupted word.

Note that by the non-singularity of the Vandermonde matrix (i.e.
polynomial interpolation), there exists a vector $\mathbf{d} =
(d_1,\ldots,d_n) \in F^n$ such that $\mathbf{d^\top \hat f}=f(0)$ for
any $f \in F[x]$ and $deg(f)<n$.

\begin{fact}
  Let $\delta'=n-\delta-1$. The duals of the codes $V^\delta, V^\delta_0$ are 
  \begin{displaymath}\
  \begin{array}{rcccl}
    W^{\delta'}&=& (V_0^\delta)^\perp &=& \set{(d_1q(1),...,d_nq(n)): \
    \deg(q) \leq \delta' }\\
    W_0^{\delta'} & =& (V^{\delta})^\perp &=& \set{(d_1q(1),...,d_nq(n)): \
    \deg(q) \leq \delta' \mbox { and }q(0)=0}
  \end{array}
\end{displaymath}
Thus the dual of a Reed-Solomon code of degree $\delta$ is another RS
code with degree $\delta'$, but where each component has been
``scaled'' according to some constant $d_i$. One can also show that
$d_i\neq 0$ for all $i$.
\end{fact}

The code $\mathcal{C}$ for parameter $\delta$ (occasionally written
$\C^\delta$) is the \css\ code obtained from codes $V=V^\delta$ and
$W=W^{\delta'}$. As mentioned before, it encodes a single qupit since
$\dim V = \delta+1$ and $\dim \wp = \delta$. Moreover, the minimum
distance of $V$ is $n-\delta$ and the minimum distance of $W$ is
$\delta+1$. Thus, for $\delta<(n-1)/2$ we get that the minimum
distance of $\mathcal{C}$ is at least $\delta+1$, and it corrects at
least $t=\delta/2$ errors.

The encoding we obtain can be described explicitly. Let $V_a^\delta =
\{ \mathbf{\hat q}:\ \deg(q) \leq \delta \mbox{ and }q(0)=a\}$. Then
for any qupit in a pure state $\ket\psi = \sum_{a \in F} \alpha_a \ket
a$, the encoded version is (ignoring normalization constants):
$$\mathcal{E}\ket\psi = \sum_a \alpha_a \mathcal{E}\ket a =  \sum_a
\alpha_a \sum_{\mathbf{v}\in V_a^\delta} \ket{\mathbf{v}} = \sum_a
\alpha_a \sum_{q: \deg(q) \leq \delta,\  q(0)=a}
\ket{q(1),...,q(n)}$$

Note that the circuit for encoding is very simple: consider the linear
map which takes the coefficients of a polynomial of degree at most
$\delta$ and maps it to the vector $q(1),...,q(n)$. Then placing
$\ket\psi$ in the position of the constant coefficient and initializing all
other coefficients to the equal superposition $\sum_a\ket a$ will yield
the output $\mathcal{E}\ket \psi$.

\paragraph{Correction, Detection and Erasures}

As mentioned above, the classical RS codes have efficient decoding
algorithms for identifying and decoding the maximum number of errors
which is information-theoretically possible, i.e. $t$ where $d=2t+1$
is the minimum distance. Consequently, so do the quantum polynomial
codes, since for \css\ codes one simply corrects errors in each of the
two bases.

They can also \emph{detect} up to $2t$ errors, at the expense of
correction. Simply measure a received codeword to see if its syndrome
is 0. If a non-zero Pauli operator of weight less than $d$ has been
applied to the word, the syndrome will be non-zero, and the error will
be detected. For an arbitrary error of weight less than $d$, the
projection of the corrupted word onto the code will be exactly the
original codeword.

\begin{rem}
  In some of our protocols, we will want to detect a large number of
  errors, but still be able to correct a small number. Suppose that we
  have identified $b$ positions which are known to be corrupted (for
  example, say they have been erased). Then the quantum polynomial
  code will be able to identify $t$ \emph{further} errors, and will
  able to correct them if there are at most $t-b$.
  
  (That is, the punctured code (i.e. restricted to the $n-b$
  non-erased positions) has distance $2t+1-b$. Given a corrupted word,
  one can tell if it is within $t-b$ of a codeword, and correct such
  errors. If it is not within distance $t-b$ of a codeword, then more
  errors occurred.  However, as long as less than $t$ errors occurred,
  the corrupted word will not be within $t-b$ of anything but the
  correct codeword, since $t+ (t-b)$ is less than the new minimum
  distance).
\end{rem}

\subsection{Sharing Quantum Secrets and (No) Cloning}
\label{sec:qss}

One of the fundamental theorems of quantum information theory is that
an arbitrary quantum state cannot be cloned. In fact, one can say
more: if there is a process with one input and two outputs, then if
one of the outputs is a copy of the input, the other output \emph{must
  be independent of the input}. We're not sure to whom this result is
attributable but it has certainly become folklore.

\begin{fact}[No cloning, folklore] 
  Let $U: {\cal H}_m\otimes {\cal H}_W \longrightarrow{\cal
    H}_A\otimes {\cal H}_B$ \textrm{(\footnote{Note that in fact the
      $mW$ system and the $AB$ system are one and the same. The two
      labelings simply reflect a different partitioning of the
      system.}) }be a unitary transformation such that for all
  $\ket{\psi} \in {\cal H}_m$:
  $$U( \ket{\psi}\otimes\ket{W}) = \ket{\psi}\otimes
  \ket{\varphi_{\ket\psi}}$$
  where $\ket{W}$ is some fixed auxiliary
  state (work bits). Then $\ket{\varphi_{\ket \psi}}$ does not depend
  on $\ket{\psi}$.
%
%
\end{fact}

An important consequence of this was first pointed out by Cleve,
Gottesman and Lo \cite{CGL99}: any quantum code is a scheme for
sharing quantum secrets: A distance $d$ code can correct $d-1$
erasures, and so access to any $n-d+1$ (uncorrupted) positions suffice
to recover the encoded state; on the other hand, that means that any
set of $d-1$ positions must reveal no information at all about the
encoded state. That is, the density matrix of any $d-1$ positions is
completely independent of the data.

Note that this phenomenon has no simple classical analogue: any
position of a classical error-correcting code will leak information
about the data unless the encoding process is randomized. This
additional step is not necessary in the quantum setting since the
randomness is somehow ``built in.''

\subsection{Tools from Fault-Tolerant Quantum Computing}
\label{sec:ftqc}

In our proposed solution, we also use techniques developed for
fault-tolerant quantum computing (\ftqc). The challenge of \ftqc\ is
to tolerate \emph{non-malicious} faults occurring within a single
computer. One assumes that at every stage in the computation, every
qubit has some probability $p$ of suffering a random error, i.e. of
becoming completely scrambled (this corresponds to the classical
notion of random bit flips occurring during a computation). Moreover,
errors are assumed to occur \emph{independently} of each other and of
the data in the computation. See \secref{prevwork} for a discussion of
the difference between \ftqc\ and \mpqc. In this section, we review a
number of useful results from \ftqc. These come from
\cite{Shor96,AB99,GC99}.

\paragraph{Universal Sets of Gates}
\label{sec:univ}

The usual technique behind fault-tolerant computing (both classical
and quantum) is to design procedures for applying one of a small
number of gates to logical (i.e. encoded) values, without having to
actually decode the values and then re-encode them. That is, given the
encoding state $\ket\psi$, we want a simple procedure which returns the
encoding of state $U\ket\psi$. 

Thus, it is useful to find a small set of gates which is
\emph{universal}, i.e which suffices to implement any desired
function\footnote{In fact, it is impossible to find a finite set which
  can implement any unitary operation perfectly. However, one can
  approximate any unitary operation on a constant number of qubits to
  accuracy $\epsilon$ using $O(poly\log \frac 1 \epsilon)$ gates from a
  ``universal'' set, i.e. one which generates a group which is dense
  in the space of all unitary operators.}. One can then simply design
fault-tolerant procedures for implementing these gates, and compose
them to obtain a fault-tolerant procedure for any particular function.

\an{Do I really want to discuss error issues more? No.}

For qupits of prime dimension $p$, Aharonov and Ben-Or \cite{AB99} showed that the
following set of gates is universal:

\begin{nar-enum}
\item \label{gate:x} Generalized NOT (a.k.a. $X$): $\forall~c\in F$, $\ket a \longmapsto \ket{a+c}$,
\item Generalized CNOT (Controlled Addition): $\ket{a,b}\longmapsto \ket{a,a+b}$,
\item Swap $\ket{a}\ket{b} \longmapsto \ket{b}\ket{a}$,
\item Multiplication gate: $0\neq c\in F$: $\ket{a}\longmapsto\ket{ac}$, 
\item \label{gate:z} Phase Shift (a.k.a. $Z$):
$\forall c\in F$ $\ket{a}\longmapsto w^{ca}\ket{a}$, 
\item Generalized Hadamard (Fourier Transform):
 $\ket{a}\longmapsto\frac{1}{\sqrt{p}}\sum_{b\in F}w^{rab}\ket{b}, \forall 0<r<p$.
\item Generalized Toffoli: $\ket{a}\ket{b}\ket{c}\longmapsto \ket{a}\ket{b}\ket{c+ab}$, 
\end{nar-enum}

Beyond these, in order to simulate arbitrary quantum circuits one
should also be able to introduce qupits in some known state (say
$\ket{0}$), as well as to discard qupits. Note that these are
sufficient for simulating measurements, since one can simply apply a
controlled-not with a state $\ket 0$ as the target and then discard
that target.

\paragraph{Transversal Operations}

Fortunately, several of these gates can be applied
\emph{transversally}, that is using only ``qupit-wise'' operations.
These are important since they correspond to operations performed
locally by the players in a multi-party protocol, if each player shares
has one component of an encoded state.

For example: in any \css\ code, the linear gate $\ket{a,b}\longmapsto
\ket{a,a+cb}$ can be applied to two encoded qupits by applying the
same gate ``qupit-wise'' to the two codewords. For any \css\ code,
the gates \ref{gate:x} through \ref{gate:z} from the set above can be
implemented transversally \cite{Shor96,AB99}.

\begin{rem}\label{rem:measure}
  Another operation which can almost be performed transversally is
  measurement in the computational basis. The encoding of a classical
  state $\ket{s}$ in a \css\ code is the equal superposition of all
  the words in some particular coset of $V_0=\wp$ within $V$. Thus,
  measuring all the qupits of the encoding of $\ket s$ will always
  yield a codeword from that coset. Similarly, measuring all the
  qupits of the encoding of $\sum_s \alpha_s \ket s$ will yield a word
  from the coset corresponding to $s$ with probability $|\alpha_s|^2$.
  This operation is not quite transversal since after the qupit-wise
  measurement, the classical information must be gathered together in
  order to extract the measurement result. Nonetheless, the quantum
  part of the processing is transversal, and this will be good enough
  for our purposes.
\end{rem}

\paragraph{Transversal Fourier Transforms and the Dual Code}

In general, applying the Fourier transform transversally to a codeword
from a \css\ code $\mathcal{C}$ does not yield a word from that code.
Instead, one obtains a word from the ``dual code'' $\TC$. If $\C$ is
defined by the classical codes $V$ and $W$, then $\TC$ is the \css\ 
code obtained using the codes $W$ and $V$. A natural choice of
encoding for the dual code yields the following relation:
$$\fn \mathcal{E}_{\C} \ket\psi = \mathcal{E}_{{\C^{\delta'}}} \paren{
  \F \ket\psi}$$
where $\mathcal{E}_\C$ and
$\mathcal{E}_{\C^{\delta'}}$ are the encoding operators for $\C$ and
${\C^{\delta'}}$ respectively.

For polynomial codes of degree $\delta$, recall that there is related
degree $\delta'=n-\delta-1$. As one can observe from the dual codes
$W^{\delta'},W_0^{\delta'}$, the dual code  $\widetilde \C^\delta$ is a
``mangled'' version of the code $\C^{\delta'}$. In fact, by scaling
each Fourier transform with the (non-zero) factor $d_i$, one obtains:
$$\F^{\bf d} \mathcal{E}_{\C^\delta} \ket\psi =
\mathcal{E}_{\C^{\delta'}} \paren{ \F \ket\psi}$$

Note that when $n$ is exactly $2\delta+1$, the codes $\C^\delta$ and
$\C^{\delta'}$ are the same, and so the Fourier transform on encoded
data can in fact be applied transversally: $\F^{\bf d}
\mathcal{E}_{\C^\delta} \ket\psi = \mathcal{E}_{\C^{\delta}} \paren{
  \F \ket\psi}$.

\paragraph{Transversal Reductions to \emph{Degree Reduction} for $\delta<n/3$}

As mentioned above, the only operations that cannot, in general, be
performed transversally on Reed-Solomon codes are the Fourier
transform and Toffoli gate. However, when $\delta$ is less than $n/3$,
\cite{AB99} reduces both of them to the problem of \emph{degree
  reduction}, which involves mapping the encoding of $\ket\psi$ under
the dual code $\C_{\delta'}$ to the encoding of $\ket\psi$ under the
original code $\C_\delta$.

\emph{For the Fourier transform}, the reduction is obvious: we showed
above that by performing (scaled) Fourier transforms transversally to
$\mathcal{E}_{\C_\delta} \ket\psi$, one obtains
$\mathcal{E}_{\C_{\delta'}} \paren{ \F \ket\psi}$. Thus, performing
degree reduction would produce $\mathcal{E}_{\C_{\delta}} \paren{ \F
  \ket\psi}$, which is the desired result.

\emph{For the Toffoli gate}, note that $\delta<n/3$ implies that
$\delta'=n-\delta-1$ is at least $2\delta$. The underlying idea is
simple: suppose we have three polynomials $p,q,r$ of degree such that
$p(0)=a, q(0)=b$ and $r(0)=c$. Take the polynomial $r'$ given by
$r'(i) = r(i) + p(i)q(i)$ for all $i=1,...,n$. First, note that if
$p,q$ have degree at most $\delta$ and $r$ has degree at most
$\delta'\geq 2\delta$, then $\deg (r') < \delta'$. Moreover, if
$p,q,r$ are \emph{random} polynomials subject to the above
constraints, then $p,q,r'$ will also form a random triple of
polynomials, which interpolate to the values $a,b,c+ab$.

To map this to a procedure for implementing the Toffoli gate, suppose
that we have the encodings of $\ket a$ and $\ket b$ using the code
${\C^{\delta}}$.  Suppose that we also have the encoding of $\ket c$
using the related code ${\C^{\delta'}}$. By applying the Toffoli gate
qupit-wise, we obtain the encoding of $c+ab$ under the related code:
$$
\E_{\C^{\delta}} \ket a \E_{\C^{\delta}} \ket b \E_{\C^{\delta'}}
\ket c \longmapsto \E_{\C^{\delta}} \ket a \E_{\C^{\delta}} \ket b
\E_{\C^{\delta'}} \ket {c+ab}$$

Thus, to implement the Toffoli gate fault-tolerantly it is sufficient
to have an implementation of the two maps $\E_{\C^{\delta}} \ket \psi
\longmapsto \E_{\C^{\delta'}} \ket \psi$ and $\E_{\C^{\delta'}} \ket
\psi \longmapsto \E_{\C^{\delta}} \ket \psi$. Note that this is
equivalent to having a procedure for just one map $\E_{\C^{\delta'}}
\ket \phi \longmapsto \E_{\C^{\delta}} \ket \phi$, since one can
simply apply the Fourier transform first and its inverse afterwards to
reverse the direction.

\paragraph{Implementing Degree Reduction}

The circuit we use for degree reduction is due to Gottesman and
Bennett \cite{Gpers} (based on \cite{GC99}), and is much more
efficient than the original one proposed in \cite{AB99}.  Begin with
the state to be transformed (call this system ${\cal H}_1$) and an
ancilla in state $\E_{\C^{\delta}} \ket{0}$ (called $\mathcal{H}_2$).

\begin{nar-enum}
\item Apply controlled addition from ${\cal H}_1$ to ${\cal H}_2$.
\item Apply the scaled Fourier transform transversally to ${\cal
    H}_1$.
\item Measure ${\cal H}_1$ in the computational basis, obtaining $b$.
\item Apply a conditional phase shift with scaling factor $-b$ to
${\cal H}_2$.
\end{nar-enum}

The effect of this on the basis state $\E_{\cal C}\ket{a}$ (for $a \in
\ze_p$) is:
\begin{eqnarray*}
  \E_{\cal C}\ket{a}\E_{\tilde {\cal C}}\ket{0} \mapsto \E_{\cal
  C}\ket{a}\E_{\tilde {\cal C}}\ket{a} &\mapsto &\sum_b \omega^{ab}
  \E_{\cal C}\ket{b}\E_{\tilde {\cal C}}\ket{a} \\
  &\mapsto &
  \omega^{ab} \E_{\tilde {\cal C}}\ket{a} \textrm{(with }b\textrm{
  known)} \mapsto \E_{\tilde {\cal C}}\ket{a}
\end{eqnarray*}

This procedure in fact works for arbitrary linear combinations
(intuitively, this is because the measurement result $b$ yields no
information about $a$). 

Note that this entire procedure can be performed transversally except
for the measurement step.  However, as noted above (\remref{measure}),
measurement requires only classical communication between the
components (namely, each component is measured and the classical
decoding algorithm for the code $V^{\delta'}$ is applied to the
result).

\afterpage{\clearpage} 

\section{Neighborhoods of Quantum Codes}
\label{sec:nbhds}

One of the ideas behind classical multi-party computing protocols is to
ensure that data is encoded in  a state that remains ``close'' to a
codeword, differing only on those positions held by cheaters, so that
error correction and detection can be used to correct any tampering,
or at least detect it and identify its origin. 

For classical codes, the notion of closeness is clear: the set of
positions on which a real word $\mathbf{v}$ differs from a codeword
provides a lot of information; in particular, the size of this set is
the Hamming distance of $\mathbf{v}$ from the code. As long as the
minimum distance of the code is at least $2t$, ensuring that
$\mathbf{v}$ differs from a codeword only on the positions held by
cheaters means that any errors introduced by cheaters will be
correctable.

Given a set $B$ of cheaters ($B\subseteq\set{1,...,n}$), we define:
\begin{eqnarray*}
  W_B &=& \set{\mathbf{v}  \ :\ \exists \mathbf{w} \in W\ \mbox{s.t.}\ \supp(\mathbf{v-w})\in
    B}\\
  &=& \set{\mathbf{v} \ :\ \exists \mathbf{w}  \in W \ \mbox{s.t. }\ \mathbf{v} \ \mbox{differs
      from}\ \mathbf{w} \ \mbox{only on positions in}\ B}
\end{eqnarray*}

Equivalently, one can define $W_B$ as the set of words obtained by
distributing a (correct) codeword  to all players, and then having all
players send their shares to some (honest) receiver/reconstructor.

\begin{rem}
  $V_B$ is a linear code, and its dual is exactly the set of words in
  $\vp$ which have support included in the complement of $C$ (say
  $A=\set{1,...,n}\setminus B$). In particular, this means that if one
  wants to measure the $V_B$-syndrome, one only needs access to
  positions in $A$.
\end{rem}

For quantum codes, the situation is more complex. For a \css\ code
$\C$, there is more than one natural definition of the neighborhood
corresponding to a set $B$ of positions. Let
$\mathcal{H}=\mathcal{H}_1 \otimes \cdots\otimes \mathcal{H}_n$ be
partitioned according to two sets $A,B$, so that
$\mathcal{H}=\mathcal{H}_A\otimes \mathcal{H}_B$. We consider three
definitions of an ``$B$-neighborhood'' of $\C$. Let $\rho$ be an
arbitrary state of $\mathcal{H}$.

For a mixed state given by density matrix $\rho'$, we say $\rho'$ is
``in'' $\C$ if all states in the mixture lie in $\C$
(no matter how the mixture is written). Algebraically, this is
given by the condition $\tr(P_\C \rho')=1$ where $P_\C$
is the projector onto the subspace $\C$.\footnote{To see why
this is the case, write $\rho'=\sum_i p_i \ket{\psi_i}\bra{\psi_i}$
with $\braket{\psi_i}{\psi_j}=\delta_{ij}$ and $\sum_i p_i=1$.
Then all the $\ket{\psi_i}$'s are in $\C$ if and only if
$\bra{\psi_i}P_\C\ket{\psi_i}=1$. Taking the trace over the
matrix $P_\C\rho$ yields 1 if and only if this condition
holds.} 

\begin{nar-enum}
\item\label{nbhd1} $\rho$ differs from a state in $\C$ only by some
  super-operator local to $B$:
  $$
  N_B(\C) = \set{\rho : \exists \rho'\ \mbox{in}\ \C,
    \exists \mathcal{O}\ \mbox{super-operator, acting only on}\ 
    \mathcal{H}_B\ \mbox{s.t.}\ \rho = \mathcal{O}(\rho') } $$
  
\item $\rho$ is cannot be distinguished from a state in $\C$ by
  looking only at positions in $A$. Algebraically, this is captured by
  requiring that the density matrix obtained by ``tracing out'' the
  positions in $B$ be the same as for some state in the code (the
  notation $ST$ stands for ``same trace''):
  $$ST_B(\C) = \set{\rho : \exists \rho'\ \mbox{in}\ \C\ \mbox{s.t.}\ 
    \tr_B(\rho) = \tr_B(\rho')}$$
  
\item \label{nbhd3} Specifically for \css\ codes, one could simply
  require that the state $\rho$ pass checks on $A$ in both bases, i.e.
  that measuring either the $V_B$-syndrome in the computational basis,
  or the $W_B$-syndrome in the Fourier basis, would yield the result
  0.  The set of states which pass this test is:
  $$\C_B = \vbq \cap \fn \wbq.$$
\end{nar-enum}

These notions form a hierarchy, namely $N_B(\C)\subseteq
ST_B(\C)\subseteq\C_B$. (The first inclusion holds since
super-operators local to $B$ do not change the density matrix of the
components in $A$. The second inclusion holds since the outcome
distribution of any tests local to $A$ is determined entirely by
$\tr_B(\rho)$.)  However, the three notions are distinct and in fact
only one of them---notion (\ref{nbhd3})---always describes a linear
subspace of $\mathcal{H}$. We discuss these three notions further in
\appref{morenbhds}.

In the analysis of quantum error-correction and fault-tolerance
protocols, it is sufficient to consider notion (\ref{nbhd1}). This
stems from two reasons. On one hand, one starts from a correctly
encoded state. On the other hand, the errors introduced by the
environment will be independent of the encoded data (and in fact they
must be for error-correction to be possible at all in that context).

In our setting, however, we cannot make such 
assumptions, since the cheaters might possess states which are entangled with
the data in the computation, and so the errors they introduce will not
be independent of that data. Instead, 
the main contribution of this paper is the construction of protocols
which guarantee conditions similar to (\ref{nbhd3}) above. In
\secref{SP}, we illustrate the ideas with a simple protocol, dubbed
\emph{subspace projection}, which is sufficient for \vqss\ and
\mpqc\ when $t<n/8$. In \secref{2ls}, we give a \vqss\ protocol
tolerating $t<n/4$, and we show that this tolerance is optimal in
\secref{imposs}. Finally, in \secref{mpc}, we show how to ensure
condition (\ref{nbhd3}) above and how the techniques from
fault-tolerant computing can then be used to achieve multi-party
computation of an arbitrary quantum circuit when $t<n/6$.

\an{Somewhere mention that in fact top-level protocol more-or-less
  guarantees condition (2), I think. This is weird...}

\subsection{Well-Definedness of Decoding for States in $\C_B$}

In this section we prove a property of $\C_B$ which will be useful in
the proof of security (and hopefully also provide some intuition for
our construction).

Suppose that the minimum distance of $\C$ is $d>2t+1$, and $B$ is
restricted in size: $|B|<t$. Then applying the usual decoding circuit
for $\C$ without knowing exactly where $B$ is yields the same result
as applying an ideal interpolation circuit which first discards
positions in $B$ and then reconstructs the logical data as if it was
handling a regular codeword. Formally, there are two natural
``reconstruction operators'' for extracting the secret out of a state
which has been shared among several players.

\begin{nar-enum}
\item ${\cal D}$ is the decoding operator for the error-correcting
  code $\mathcal{C}$. For any operator $E_j$ of weight less than $t$
  and for any state $\ket{\bar \phi}$ in $\mathcal{C}$, we have
  $\mathcal{D}E_j\ket{\phi} = \ket{\phi} \otimes \ket{j}$ (i.e. the
  error is not only corrected but also identified). It will then
  discard the system containing the syndrome information $\ket{j}$.
\item $\ri$ is the ``ideal recovery operator'', defined by identifying
  the set $B$ of cheaters and applying the simple interpolation
  circuit to a set of $n-2t$ good players' positions.
\end{nar-enum}

\begin{prop}\label{prop:l1-well-defined}
  For any state $\rho$ in $\mathcal{C}_B$ where $|B|\leq t$, the state
  $\ri (\rho)$ is well-defined and is equal to $\mathcal{D}(\rho)$.
\end{prop}

We give the proof of this below. For now, note that
\propref{l1-well-defined} means that no changes made only to the
components in $B$---no matter how they might be made to interact with
outside systems entangled with the data---will change the
reconstructed state.

In order to prove \propref{l1-well-defined}, we characterize
$\C_B$ algebraically:

\begin{lemma}\label{lem:charact}
  Suppose that $\rho$ has fidelity 1 to $\C_B=\vbq \cap \fn
  \wbq$. Then we can write
  \begin{eqnarray*}
    &\rho=\sum_i p_i\ket{\psi_i}\bra{\psi_i}&\\ &\ket{\psi_i} = \sum_j
    c_{ij}E_j\ket{\phi_{ij}}&
  \end{eqnarray*}
  where $E_j$ are Pauli operators on $B$ and $\ket{\phi_{ij}}\in {\cal
  C}$.
\end{lemma}

Recall that given a state $\rho$, testing if $\rho$ is in $\vbq$ is
easily described: For each element of (a basis of) the dual space
$\vbp$, we measure the corresponding linear combination of the qupits
of $\rho$ in the computational basis, and check that it is 0. Recall
that the vectors of the dual space $\vbq$ have support only on $A$
(since arbitrary changes to positions in $B$ should not affect whether
or not a word is in $V_B$), and so one need not have access to the
components in $A$ in order to perform the measurement. Similarly, to
check if $\rho$ is in $\fn\wbq$, we rotate into the Fourier basis and
measure the linear combinations corresponding to a basis of $\wbp$.

Note that since $\vbp \subseteq \vp$ and $\wbp\subseteq \wp$, and
since measuring the $V$-syndrome in the computational basis commutes
with measuring the $W$-syndrome in the Fourier basis, we know that the
following four measurements commute:

\begin{nar-enum}
\item $V_B$-syndrome in the computational basis
\item $V$-syndrome in the computational basis
\item $W_B$-syndrome in the Fourier basis
\item $W$-syndrome in the Fourier basis
\end{nar-enum}




\begin{proofof}{\lemref{charact}}
  As was just mentioned, to check if $\rho$ is in $\C_B$, we measure
  the $V_B$-syndrome in the computational basis and the $W_B$-syndrome
  in the Fourier basis.  But by the remarks above, the distribution on
  this outcome measurement will not change if we first measure the $V$
  and $W$ syndromes, i.e.  if we first make a measurement which
  projects $\rho$ into one of the subspaces $E_j\C$ (i.e. $\rho$ maps
  to $\rho'=P_j \rho P_j$ with probability $\tr\paren{P_j \rho}$,
  where $P_j$ is the projector for the space $E_j\mathcal{C}$).
  
  The new state $\rho'$ lies completely in one of the spaces $E_j$.
  However, $E_j\C$ is either contained in $\C_B$ (if there is an
  operator equivalent to $E_j$ which acts only on $B$) or
  \emph{orthogonal} to $\C_B$ (if no such operator exists).

  Thus, for $\rho$ to have fidelity 1 with $\C_B$, it must be
  that $\tr\paren{P_j\rho}=0$ for all $E_j$ which act on more than
  $B$. Hence $\rho$ is a mixture of states $\ket{\psi_i}$ each of
  which is a linear combination of elements of the spaces
  $\set{E_j\C}$, where $E_j$ acts only on $B$.
\end{proofof}

\begin{proofof}{\propref{l1-well-defined}}
%
  Consider a particular basis state $E_j\E \ket a$. The decoding
  operator $\mathcal{D}$ will produce the state $\ket a\ket j$, since
  errors of weight at most $t$ can be identified uniquely.  The ideal
  operator $\ri$ will extract the encoded state $\ket{a}$. Without
  loss of generality, the ideal recovery operator will replace $\ket
  a$ with $\ket 0$, the final output $\ket a \otimes E_j\E\ket{0}$.
  
  In both cases, the output can be written as $\ket{a}$ tensored with
  some ancilla whose state depends only on the syndrome $j$ (and which
  identifies $j$ uniquely).  Once that state is traced out, the
  outputs of both operators will be identical. Another way to see this
  is that the ideal operator can simulate the real operator: one can
  go from the output of the ideal operator to that of the real
  operator by applying a transformation which only affects the
  ancilla. For a state $\rho$ expressed as in \lemref{charact}, the
  final outcome will be $\rho' = \sum_{ij} p_i |c_{ij}|^2
  \proj{\phi_{ij}}$.
\end{proofof}

\afterpage{\clearpage} 

\chapter{Distributed Protocols for Quantum Computers}
\label{cha:protocols}


\section{Subspace Projection}
\label{sec:SP}

Before presenting the main \vqss\ protocol, we describe a protocol for a
simpler task that we call {\em subspace projection}, which illustrates
the key ideas in the \vqss\ protocol. Namely, we first modify a
classical protocol of \cite{CCD88} so that the dealer does not have to
remember the random bits he used in sharing his secret.  Second, we
apply this protocol both in the computational and Fourier bases. We
use a ``quantum-to-classical'' argument to show that this garantees
that the joint state shared by the players satisfies condition (3)
from the discussion on neighborhoods, i.e. that the joint state passes
certain local checks in both bases.

Recall that for any given system of $n$ qupits and any linear subspace
$W$ of $F^n=\ze_p^n$, we define
$$W^{(q)} = \mathrm{span}\{\ket{{\bf w}}:\ {\bf w} \in W\}.$$
For this
protocol, $W$ can be any code with minimum distance $2t+1$ and an
efficient decoding algorithm. However, for concreteness, let $W$ be
the RS code $V^\delta$, where $n=4t+1$ and $\delta=2t$.

Let $\mathcal{H}_0,\ldots,\mathcal{H}_{k}$ be separate quantum systems
consisting of $n$ qupits each, and let $\mathcal{H} = \mathcal{H}_0
\otimes \cdots \otimes \mathcal{H}_{k}$. Say the dealer prepares
$\mathcal{H}$ in some state and gives the $i$th qupit of each
subsystem $\mathcal{H}_j$ to player $i$.  He wants to prove to the
group that in fact the fidelity of $\mathcal{H}_0$ to the space
$W^{(q)}$ is close to 1 \footnote{It would be desirable to be able to
  prove that the fidelity is in fact exactly 1. This remains an
  interesting open question. This corresponds to the classical
  difference between zero-error and small-error protocols.}, without
revealing any information beyond that to the other players.  What we
achieve in this first step is not quite that strong: at the end of the
protocol, there will be a publicly known set $B$ of ``apparent
cheaters'' such that the shares of the honest players not in $B$ will
all agree with $W$ in the computational basis, i.e. will have high
fidelity to the space $W_{B\cup C}^{(q)}$.

We obtain a ``cut-and-choose'' protocol, also similar to the ``random
hashing'' technique used in purification protocols (\protref{sp},
\figref{prot-sp}). Note that \textsc{vss} and broadcast of classical
data are not a problem since $t<\frac{n}{4}<\frac{n}{3}$
(\cite{BGW88,CCD88,Lynch}).

\begin{figure}[h]
  \begin{center}
    \fbox{\parbox[t]{\textwidth}{{\small
\begin{protocol}{Subspace projection}\label{prot:sp}{~}
  \begin{nar-enum}
  \item \textbf{Sharing} The dealer $D$ prepares $\mathcal{H}_0$ as any
  state (pure or mixed) in $W^{(q)}$ and distributes it to the
  players. He then prepares $\mathcal{H}_1,\ldots,\mathcal{H}_k$ in
  the equal superposition of $\sum_{{\bf w} \in W} \ket{{\bf w}}$, and
  distributes those to the players also.

\item \textbf{Verification} Using classical \textsc{vss}, every player
  commits to $k$ field elements picked uniformly at random. These
  commitments are then opened and their sum is taken to obtain $k$
  field elements $b_1,\ldots,b_k$ (these are completely unpredictable
  to the dealer, even if he is cheating).

\item \label{step:sp-add} For $\ell=1,\ldots,k$, players apply the
  linear operation $(x,y)\mapsto(x,y+b_\ell x)$ to the subsystems
  ${\cal H}_0$ and ${\cal H}_\ell$.  All players then measure their
  shares of ${\cal H}_1,\ldots,{\cal H}_k$ in the computational basis
  and broadcast the result.
  
\item Each of the broadcasted words ${\bf w}_1,\ldots,{\bf w}_k$ is
  decoded using classical error-correction of the code $W$: for each
  ${\bf w}_\ell$, we obtain either that it was at distance more than
  $t$ from a word in $W$ or we obtain an error vector with support
  $B_\ell \in \set{1,...,n}$ of size less than $t$ on which ${\bf
    w}_\ell$ differs from a word in $W$.

  The dealer is rejected if any of the broadcasted words was at
  distance more than $t$ or if $B=\bigcup_{\ell=1}^{k} B_\ell$ has size
  greater than $t$. Otherwise, the dealer is accepted.
\end{nar-enum}
\end{protocol}
}}}\caption{\protref{sp} (Subspace Projection)}
\label{fig:prot-sp}
\end{center}
\end{figure}

\subsection{Completeness}

\begin{lemma}
  When the dealer $D$ is honest, he will pass the protocol. Moreover,
  we will have $B\subseteq C$, i.e. only real cheaters will be accused
  of cheating.
\end{lemma}

\begin{proof}
  If the dealer is honest, he will use some ${\cal H}_0$ in $W^{(q)}$
  and will have all ${\cal H}_\ell$'s in state $\sum_{{\bf w} \in W}
  \ket{{\bf w}}$. Consider some round $\ell$. Now no matter what the
  value of $b_\ell$ is, applying $(c\mbox{-}X^{b_\ell})$ to \emph{all} of the
  shares is equivalent to the identity on ${\cal H}_0\otimes {\cal
    H}_\ell$, since for all $\mathbf{v}\in W$, we have:
  $$(c\mbox{-}X^{b_\ell})\ket{\mathbf{v}}\sum_{\mathbf{w} \in
    W}\ket{\mathbf{w}} = \ket{\mathbf{v}}\sum_\mathbf{w}
  \ket{\mathbf{w}+b_\ell \mathbf{v}}= \ket{\mathbf{v}}\sum_\mathbf{w}
  \ket{\mathbf{w}}$$

  Of course, in the protocol we can only guarantee that honest players
  will apply $(c_X^{b_\ell})$ to their shares of ${\cal H}_0$ and
  ${\cal H}_\ell$. Nonetheless, the result is the same as applying the
  identity to the honest players' shares. Consequently, the values
  broadcast at \stepref{sp-add} by the honest players will all
  be consistent with some $\mathbf{w} \in W$. Since we've assumed that the
  distance of the code $W$ is at least $2t+1$, any false values
  broadcast by cheaters will be identified as such. Thus, the set $B$
  will only contain cheaters, and the dealer will pass the protocol.
  Moreover, the honest players' shares of ${\cal H}_0$ will also be
  preserved, so ${\cal H}_0$ will remain in $W_C^{(q)}$.
\end{proof}

\subsection{Soundness}
\label{sec:sp-sound}

\begin{lemma}
  \label{lem:SP}
  Let $\tilde B = B \cup C$. At the end of the protocol above, the
  fidelity of the system to the statement ``either ${\cal H}_0$ is in
  $(W_{\tilde B})^{(q)}$ or the dealer has been rejected'' is
  exponentially close to 1 in $k$.
\end{lemma}

To prove this, we will employ a ``quantum to classical'' reduction, as
in \cite{LC99kd}.

\begin{lemma}
  \label{lem:q2c}
  Consider the subspace projection protocol above. Then the
  behavior of the protocol is the same in each of the two following experiments:
  \begin{nar-desc}
  \item[Experiment 1] at the end of the \emph{whole protocol}, all
    honest players measure their shares of ${\cal H}_0$ in the
    computational basis, \emph{or}
  \item[Experiment 2] at the end of the \emph{sharing phase}, all
    honest players measure their shares of ${\cal H}_0$ and ${\cal
      H}_1$ in the computational basis, and then run the verification
    phase.
  \end{nar-desc}
  Moreover, the distribution on the results of the measurement of
  ${\cal H}_0$ is the same in both cases.
\end{lemma}

\begin{proof}\an{Is this proof clear enough?}
  The actions of the honest players on their shares in the original
  protocol can be seen as the composition of $k$ super-operators, each
  of which is comprised of two operations: a controlled-addition gate
  from ${\cal H}_0$ to ${\cal H}_\ell$ followed by measurement of
  ${\cal H}_\ell$. Denote the controlled-addition gate by
  $(c\mbox{-}X^b)_\ell$, where $b$ is the scaling factor for the
  controlled-addition. Second, denote measurement of ${\cal H}_\ell$
  in the computational basis by ${\cal M}_\ell$.
  
  Consider what happens in the $\lth$ verification step in Experiment
  1. Because the controlled-addition gate is a permutation of the
  basis states of the computational basis, measuring the systems in
  that basis \emph{before} the gate is applied will not change the
  outcome of measurements made \emph{after} the gate is applied. Thus
  we can write ${\cal M}_\ell {\cal M}_0 (c\mbox{-}X^{b_\ell})_\ell ={\cal
    M}_\ell {\cal M}_0(c\mbox{-}X^{b_\ell})_\ell{\cal M}_\ell{\cal M}_0$, and
  the distribution of the measurements made after the gate is applied
  will not change.
  
  But now notice that measuring the system ${\cal M}_0$ afterwards is
  completely redundant. Because the controlled-addition gate does not
  change the first component of any basis vectors, measuring ${\cal
    M}_0$ after the application of the gate will yield the same result
  as measuring it before. Hence, we can write ${\cal M}_\ell {\cal
    M}_0(c\mbox{-}X^{b_\ell})_\ell ={\cal M}_\ell (c\mbox{-}X^{b_\ell})_\ell {\cal
    M}_\ell{\cal M}_0$. However, this is exactly the sequence of
  operations performed by honest players in Experiment 2: first they
  measure both systems, then apply the addition gate and measure the
  target.

  Thus, the measurement outcomes will be the same in both experiments
  will be the same. Moreover, the cheaters can see no difference
  between the two experiments, and so their behavior will not change.
\end{proof}

In other words, we can imagine two situations. In the first one, just
after the sharing phase of the protocol, an outsider comes in and
secretly measures honest players' shares of
$\mathcal{H}_0,\ldots,{\cal H}_k$ in the computational basis. In the
second, the outsider performs this secret measurement \emph{after} the
protocol is completed. The statement is that exactly when he makes his
measurement will not change the behavior of the protocol.

But recall that our original statement of the correctness of the
protocol is that, at the end of the protocol, either the dealer has
been caught or the shares of players are in $W_{\tilde B}^{(q)}$.
Since fidelity to $W_{\tilde B}^{(q)}$ is the same as the probability
that measuring in the computational basis gives a word in $W_{\tilde
  B}$ (i.e. agrees with $W$ when truncated to positions neither in $B$
nor $C$), \lemref{q2c} allows us to restrict ourselves to thinking
about situations in which the shares of the systems ${\cal
  H}_0,\ldots,{\cal H}_k$ sent to honest players were in fact
classical states.

Now consider the classical protocol corresponding to the subspace
projection protocol: the dealer distributes $k+1$ codewords ${\bf
  w}_0,\ldots,{\bf w}_k$. At each step, a random multiple of ${\bf
  w}_0$ is added to one of the other codewords and the result is
broadcast. At the end, players compute $B$ as above and decide whether
or not to reject the dealer. (This is the blob protocol of
\cite{CCD88}, modified so as not to require the involvement of the
dealer beyond the sharing stage).

\begin{lemma}[Soundness of Modified Blobs from \cite{CCD88}]
  \label{lem:cSP}
  At the end of classical protocol, let $A$ be the set of honest
  players not in $B$. The event ``either the players in $A$ have
  consistent shares or the dealer was caught'' occurs with probability
  at least $1-2^{n-k}$, \emph{even when the adversary is adaptive}.
\end{lemma}
\begin{proof}
  Note that this statement is the same as Pr(the players in $A$ do not
  have consistent shares \emph{and} the dealer was not caught)$ < 2
  ^{n-k}$.

  Recall that the adversary is adaptive, and can choose which set of
  players to corrupt on the fly.  Nonetheless, the adversary's
  strategy can be reduced to choosing the set $A$ of players who will be
  neither corrupted ($\in C$) nor accused ($\in B$), but such that
  ${\bf w}_0$ is not consistent on $A$, while the broadcast vectors
  ${\bf w}_\ell + b_\ell{\bf w}_0$ are all consistent.

  Fix any particular set $A$. If the shares of ${\bf w}_0$ are not
  consistent on $A$, then there is at most a single value $b_\ell \in
  F$ such that the shares of ${\bf w}_\ell + b_\ell {\bf w}_0$
  broadcast by players in $A$ will be consistent, since the set of
  consistent vectors is a subspace. Thus, the probability of the
  dealer passing the tests with that set $A$ is at most
  $\frac{1}{|F|^k}$. Overall, there are at most $2^n$ choices for the
  subset $A$, and so the adversary's total probability of being able
  to find a subset $A$ of honest players for which cheating is
  possible is bounded above by $\frac{2^n}{|F|^k} \leq 2^{n-k}$.
\end{proof}

This completes the proof of \lemref{SP}.

\begin{rem}\label{rem:adapt-sound}
  As mentioned in the Definitions, we do not handle adaptive
  adversaries explicitly in this thesis. However, we believe that our
  protocols are secure against an adaptive adversary, and the previous
  proof gives some flavor of how the classical arguments can be used.
  In this case, a union bound argument was sufficient. For proving the
  security of the quantum protocols, a more sophisticated version of
  the quantum-to-classical reduction above (\lemref{q2c}) would be
  necessary (and, we believe, sufficient).
\end{rem}

\subsection{Dual Subspace Projection}
\label{sec:dualSP}

Consider a ``dual'' version of the subspace projection protocol above.
It is the same as the original protocol, with three changes:
\begin{nar-enum}
\item Before proceeding to the verification phase all players apply
  the Fourier transform to all their shares.
\item At the end all players apply the inverse Fourier transform to
  their shares of ${\cal H}_0$.
\item (When $D$ is honest) $D$ prepares the ancillas
  $\mathcal{H}_1,...,\mathcal{H}_k$ as a superposition over all words
  from the dual code $W^\perp$ (i.e.
  $\sum_{\mathbf{w}\in{}W^\perp}\ket{\mathbf{w}}$).
\end{nar-enum}

Now the state $\sum_{\mathbf{w}\in W^\perp}\ket{\mathbf{w}}$ is the
image of $\sum_{\mathbf{w}\in W}\ket{\mathbf{w}}$ under transversal
Fourier transforms. Thus, we can use the same analysis as in the
previous section. At the end of this ``dual'' protocol, the fidelity
of the system to the statement ``either the dealer is caught or ${\cal
  H}_0$ is in the space ${\cal F}^{\otimes n} W_{\tilde
    B}^{(q)}$'' is high.

But recall that conjugating by Fourier rotations maps linear gates to
linear gates (see \secref{prelim}). In particular, controlled addition
gates simply have their direction reversed, i.e. source and target are
swapped. Thus, the modifications to the original subspace projection
protocol can be restated as follows:
\begin{nar-enum}
\item the controlled addition gates are performed \emph{from} ${\cal
    H}_\ell$ \emph{to} ${\cal H}_0$;
\item the measurements are made in the rotated (Fourier) basis;
\item (When $D$ is honest) $D$ prepares the ancillas
  $\mathcal{H}_1,...,\mathcal{H}_k$ as a superposition over all words
  from the dual code $W^\perp$ (i.e.
  $\sum_{\mathbf{w}\in{}W^\perp}\ket{\mathbf{w}}$).
\end{nar-enum}

\paragraph{``One-Level'' Sharing and \vqss\ for $t<n/8$}

Now suppose that there is some other code $V$ such that before the
protocol begins, all the systems ${\cal H}_0,\ldots,{\cal H}_k$ are in
$V_{\tilde B}^{(q)}$. Then that property will not be affected by the
protocol since the addition gates will not affect it. Thus, at the end
of the protocol the shares of ${\cal H}_0$ would be in
$\C_{\tilde{B}}=V_{\tilde B}^{(q)} \cap \fn W_{\tilde B}^{(q)}$.

This leads to a first pass at a quantum sharing protocol: Have the
dealer distribute $k+1$ groups of $k+1$ systems. In each group, use
$k$ of the systems to prove that the remaining system lies in
$V_{\tilde B}^{(q)}$ using the subspace projection protocol. Next,
take the $k+1$ resulting systems, and use $k$ of them to prove that
one of them is also in $\fn W_{\tilde B}^{(q)}$ using the ``dual''
protocol.

Intuitively, this combination of the subspace projection protocol and
the dual protocol achieves \vqss\ when $t<n/8$: since both the sets of
apparent cheaters and of real cheaters have size at most $t$, the
protocol allows the dealer to guarantee that the shared state is in
$\C_{\tilde B}$ where $|\tilde{B}|< n/4$. Since the decoding operator
is well-defined on such states (\propref{l1-well-defined}), the dealer
is essentially committed to a unique value regardless of any changes
the players make subsequently.

In the next section, we extend the ideas of this section, combining
them with the classical \cvss\ protocol of \cite{CCD88} to obtain a
\vqss\ protocol which is secure for $t<n/4$. We also show how to prove
equivalence of that protocol to the ideal-model protocol of
\secref{vqss-def}.


\afterpage{\clearpage} 
\section{{\sc Vqss} Protocol: Two-Level Quantum Sharing}
\label{sec:2ls}

In this section we define a two-tiered protocol for \vqss. It is based
on the \vqss\ protocols of \cite{CCD88} as well as on the literature on
quantum fault-tolerance and error-correction, most notably on
\cite{AB99}.

We first define the classical notion of ``correctness'' of a sharing
used in \cite{CCD88}, and give a modified version of the \cite{CCD88}
\cvss\ protocol that does not require the dealer's participation. We
then describe our \vqss\ protocol (\secref{2ls}) and prove its
security (\secref{l2-sound}--\secref{l2-sim}). Finally, we state the
round and communication complexity of our protocol
(\secref{l2-complexity}) and some additional useful properties of the
sharings it generates (\secref{l2p}).

\subsection{Sharing Shares: $\tgood$ Trees}
\label{sec:hier}

In the \cvss\ protocol of \cite{CCD88}, the dealer $D$ takes his
secret, splits it into $n$ shares and gives the $\ith$ component to
player $i$. Player $i$ then shares this secret by splitting it into
$n$ shares and giving player $j$ the $\jth$ share to player $j$. Thus,
there are $n^2$ total shares, which can be thought of as the leaves of
a tree with depth 2 and fan-out $n$: each leaf is a share; the $\ith$
branch corresponds to the shares created by player $i$, and the root
corresponds to the initial shares created by the dealer. Thus player
$j$ holds the $\jth$ leaf in each branch of this tree.

We will run a cut-and-choose protocol similar to the subspace
projection protocol above, in order to guarantee some kind of
consistency of the distributed shares.

During the protocol we accumulate $n+1$ sets of apparent cheaters: one
set $B$ for the dealer (this corresponds to a set of branches
emanating from the root), and one set $B_i$ for each player $i$ (this
corresponds to a subset of the leaves in branch $i$).  These sets all
have size at most $t$.

N.B.: Since the dealer is one of the players in the protocol, we can
in fact identify $B$ with $B_i$, where the dealer is player $i$.
However, by ignoring this fact we lose no correctness and gain some
simplicity in the exposition and security proof of the protocol.

At the end of the protocol, we want to guarantee certain invariants:
\begin{definition}[$\tgood$ trees]\label{def:tgood} 
  We say a tree of $n^2$ field elements is $\tgood$ with respect to
  the code $V$ and the sets $B,B_1,...,B_n$ if:
  \begin{nar-enum}
  \item For each $i \not \in C$ (corresponding to an honest player),
    we have $B_i\subseteq C$, i.e. all apparent cheaters are really
    cheaters.
  \item For each branch $i \not \in B$, the shares held by the honest
    players \emph{not in} $B_i$ should all be consistent with some
    polynomial of degree $\leq d$, i.e. with some codeword in $V$.
    That is, the vector of all shares should be in $V_{B_i \cup C}$,
    where $C$ is the set of cheating players.

    N.B.: Because there are at most $t$ players in $B_i$ and at most
    $t$ cheaters, there are at least $d+1\leq n-2t$ honest players
    remaining, and so the polynomial above is uniquely defined. This
    guarantees that for each branch $i \not \in B$, there is a unique
    value $a_i\in F$ which is obtained by interpolating the shares of
    the honest players not in $B_i$.

  \item For $i\not \in B$, the values $a_i$ defined by the previous
    property are all consistent with a codeword of $V$ (i.e. the
    vector $(a_1,...,a_n)$ is in $V_B$).
  \end{nar-enum}
  We will abbreviate this as $\tgood_V$, when the sets
  $B,B_1,...,B_n$ are clear from the context.
\end{definition}

Why is this a useful property to guarantee? It turns out
that this ensures the soundness of a sharing protocol.
Suppose that all players broadcast their shares of a given $\tgood$
tree. Call the vector of shares in the $\ith$ branch $\mathbf{v}_i$,
so that player $j$ holds the values $\mathbf{v}_i(j)$ for all $i$.
Consider the reconstruction procedure $\mathsf{Recover}$
(\figref{alg-recover}).

\begin{figure}[hb!]
  \begin{center}
    \fbox{\parbox[t]{\textwidth}{
        \begin{algm}\label{alg:2goodrecover}$\mathsf{Recover}(T,V,B,B_1,...,B_n)$

          \noindent Input: a tree $T$ which is $\tgood$ with respect to the
          code $V$ and the sets $B,B_1,...,B_n$.\\
          Output: $a \in F$
          \begin{nar-enum}
          \item For each branch $i\not \in B$: Let $b=|B_i|$. If $i$
            is honest, then we expect the truncated word
            $\mathbf{v}_i|_{\bar B_i}$ to be within distance $t-b$ of
            a codeword in the truncated code $V|_{\bar B_i}$.  Now
            this truncated code has distance $2t+1-b$: it can detect
            up to $t$ errors and correct them when there are at most
            $t-b$ of them.
            
            If the truncated word $\mathbf{v}_i |_{\bar B_i}$ is at
            distance at most $t-b$ from a real codeword, then correct
            the error and let $a_i$ be the interpolated value for that
            codeword.  Otherwise output a null value $a_i = \perp$.
            
          \item Take any set of $d+1$ indices $i$ such that $i\not \in
            B$ and $a_i \neq \perp$. Find the unique polynomial $p$
            such that $p(i)=a_i$.  Output $a=p(0)$ as the
            reconstructed secret.
          \end{nar-enum}
        \end{algm}
      }}
    \caption{\algref{2goodrecover} (Reconstruction for a $\tgood$
      tree)}
    \label{fig:alg-recover}
  \end{center}
\end{figure}

\begin{lemma}\label{lem:2good}
  Suppose that a sharing is $\tgood$. If all players broadcast their
  shares, then the same value $a$ will always be reconstructed for the
  root of the tree (i.e. regardless of the values broadcast by the
  cheaters).
\end{lemma}

We omit this proof here, since it is essentially re-proven in our
analysis of the quantum protocol (see \lemref{l2-well-defined}).  We
note that the protocols (and proofs) of \cite{CCD88} used this lemma
implicitly, but did not use the recovery algorithm as stated here.
Instead, they required players to remember what shares they had
distributed to other players.

\subsection{Classical {\sc Vss}}
\label{sec:classic}

Based on the discussion of the previous section, we give a modified
version of the \cvss\ protocol of \cite{CCD88}. The main difference is
that the original protocol required a dealer to remember the values of
the shares sent in the first phase, and cooperate later on during the
verification phase. However, this does not generalize well to the
quantum world, and so we compensate by exploiting the efficient
decodability of Reed-Solomon codes. The protocol is given in
\figref{vss}. Note that as before, the error-correcting code we use is
$V^{\delta}$, where $n=4t+1$ and $\delta=2t$.

\begin{rem}
  In the description of the protocol (and subsequent protocols), we
  assume for simplicity that there is a source of public randomness.
  This is not a problem in our setting as good random bits can be
  generated using classical \cvss\ protocols, and it simplifies the
  analysis of the protocols. However, it is not necessary (and is not
  made in \cite{CCD88,RB89}).  See \secref{l2-complexity} for further
  discussion.
\end{rem}

\begin{figure}[t!]
  \begin{center}
    \fbox{\parbox[t]{\textwidth}{{\small
          \begin{protocol}{Modified Classical \cvss\ from \cite{CCD88}}\label{prot:vss} 
            The dealer $D$ has input $a\in F$.
  \begin{itemize}
  \item \textbf{Sharing:}
    \begin{nar-enum}
    \item $D$ picks a random codeword $\mathbf{v}_0\in V$ such that
      $\mathbf{v}_0$ interpolates to $a$. $D$ also picks $k$ random
      codewords $\mathbf{v}_1,...,\mathbf{v}_k\in V$ (i.e. $k$
      sharings of random values).
    \item $D$ gives player $i$ the $\ith$ component of each of these
      vectors: $\mathbf{v}_\ell(i)$ for $\ell =0,...,k$.
    \item Player $i$ shares each of these values with random vectors
      $\mathbf{v}_{0,i},...,\mathbf{v}_{k,i}$ which interpolate to
      $\mathbf{v}_0(i),...,\mathbf{v}_k(i)$, respectively. He sends
      the values $\mathbf{v}_{\ell,i}(j)$ to player $j$ (for
      $\ell=0,...,k$).
    \end{nar-enum}
  \item \textbf{Verification:} Get $k$ previously unknown public
    random values $b_1,...,b_k$. For $\ell=1,...,k$:
    \begin{nar-enum}
    \item For all $i$, player $j$ broadcasts
      $\mathbf{v}_{\ell,i}(j)+b_\ell \mathbf{v}_{0,i}(j)$.\\(i.e.
      player $j$ broadcasts his share of $\mathbf{v}_{\ell,i}+b_\ell
      \mathbf{v}_{0,i}$).
    \item \an{This needs more precision.} For each $i\in
      \set{1,...,n}$, players update the set $B_i$ based on the
      broadcast values, as in the subspace projection protocol. If
      there are too many errors, then they add $i$ to the global set
      $B$.
    \item Furthermore, players do the same at the branch level: for
      all $i \not \in B$, there is an interpolated value $a_i$ which
      corresponds to the decoded codeword from the previous step.
      Players also decode the codeword $(a_1,...,a_n)$ and update $B$
      accordingly (i.e. by adding any positions where errors occur to
      $B$).
    \end{nar-enum}
    
  \item The dealer is disqualified if $B$ is ever larger than $t$.
    
  \item If the dealer passes, the values $\mathbf{v}_{0,i}(j)$ are taken to be
    the shares of the dealer's secret.

\item \textbf{Reconstruction:}
  \begin{nar-enum}
  \item Player $j$ broadcasts his shares $\mathbf{v}_{0,i}(j)$ for all $i$.
  \item Let $T$ be the tree defined by these values. All players
    output the value given by $\mathsf{Recover}(T,V,B,B_1,...,B_n)$.
  \end{nar-enum}
\end{itemize}
\end{protocol}
}}} 
\caption{\protref{vss} (Modified \cvss\ protocol from
  \cite{CCD88})}
\label{fig:vss}
\end{center}
\end{figure}

\afterpage{\clearpage}

\an{Emphasize that additional power comes from resharing done by all
  players.}

The correctness and soundness of this protocol are stated here. They
follow from the properties of $\tgood$ trees and from cut-and-choose
analysis.

\begin{fact}
  If $D$ is honest, he will pass the protocol with probability 1, and
  the shares $\mathbf{v}_{0,i}(j)$ will form a $\tgood$ tree which interpolates
  to the original input $a$.
\end{fact}

\begin{fact}
  With probability $1-2^{\Omega(k)}$, either the shares $\mathbf{v}_{0,i}(j)$
  form a $\tgood$ tree or the dealer is caught during the protocol.
\end{fact}

\subsection{{\sc Vqss}\ Protocol}
\label{sec:vqss}

Given the previous protocol, and the observation that Subspace
Projection can work simultaneously in both bases (\secref{dualSP}), it
is natural to attempt to run the classical \cvss\ to check for errors
in both bases. The resulting protocol is described in \figref{vqss}
(Sharing Phase) and \figref{qr} (Reconstruction Phase).  Intuitively,
it guarantees that a tree of quantum shares would yield a $\tgood$
tree of classical values if measured in either the computational basis
or the Fourier basis.  Note that we use the codes
$V=V^{\delta}=V^{\delta'}$ and $W=W^{\delta}=W^{\delta'}$ (again with
$n=4t+1,\delta=\delta'=2t$), although there is in fact no need to do
this: the protocol will work for any \css\ code with distance at least
$2t+1$, so long as the code is efficiently decodable.

\an{Describe \vqss\ protocol in English.}

\begin{figure}[t!]
  \begin{center}
    \fbox{\parbox[t]{\textwidth}{{\small
\begin{protocol}{\vqss---Sharing Phase} \label{prot:vqss} 
  Dealer $D$ gets as input a quantum system $S$ to share. 
  \begin{nar-item}
  \item \textbf{Sharing:}
    \begin{nar-enum}
    \item The dealer $D$ prepares $(k+1)^2$ systems of $n$ qupits
      each, called $S_{\ell,m}$ (for $\ell=0,...,k$ and $m=0,...,k$):
      \begin{nar-enum}
      \item Encodes $S$ using ${\cal C}$ in $S_{0,0}$.
      \item Prepares $k$ systems $S_{0,1},...,S_{0,k}$ in the state
        $\sum_{a \in F} \E_\C \ket{a} = \sum_{v \in V} \ket{v}$.
      \item Prepares $k(k+1)$ systems $S_{\ell,m}$, for $\ell=1,...,k$
        and $m=0,...,k$, each in the state $\ket{\bar 0}= \sum_{v \in
          V_0} \ket{v}$.
      \item For each of the $(k+1)^2$ systems $S_{\ell,m}$, $D$ sends
        the $\ith$ component (denoted $S_{\ell,m}^{(i)}$) to player $i$.
      \end{nar-enum}
    \item Each player $i$, for each $\ell,m=0,...k$:
      \begin{nar-enum}
      \item Encodes the received system $S_{\ell,m}^{(i)}$ using
        ${\cal C}$ into an $n$ qupit system $S_{\ell,m,i}$.
      \item Sends the $j$-th component $S_{\ell,m,i}^{(j)}$ to player
        $j$.
      \end{nar-enum}
    \end{nar-enum}
  \item \textbf{Verification:}
    \begin{nar-enum}
    \item \label{step:l2-vcomp} Get public random values $b_1,...,b_k
      \in_R F$. For each $\ell=0,...,k$, $m=1,...,k$, each player $j$:
      \begin{nar-enum}
      \item Applies the controlled-addition gate $(c\mbox{-}X^{b_j})$
        to his shares of the systems $S_{\ell,0,i}$ and
        $S_{\ell,m,i}$.
      \item Measures his share of $S_{\ell,m,i}$ and broadcasts the
        result\\ (i.e. each player broadcasts $k(k+1)n$ values).
      \item Updates sets $B$ and $B_1,...,B_n$ as in the classical \cvss\
        protocol.
      \end{nar-enum}
    \item \label{step:l2-rotate} All players apply the Fourier
      transform $\F$ to their shares.
    \item \label{step:l2-vfour} Get public random values
      $b_1',...,b_k' \in_R F$. For $\ell=1,...,k$, each player $j$:
      \begin{nar-enum}
      \item Applies the controlled-addition gate $(c\mbox{-}X^{b_j'})$
        to his shares of the systems $S_{0,0,i}$ and $S_{\ell,0,i}$.
      \item Measures his share of $S_{\ell,0,i}$ and broadcasts the
        result\\ (i.e. each player broadcasts $kn$ values).
      \item Updates sets $B$ and $B_1,...,B_n$ as in classical \cvss\
        protocol. Note that \emph{for all} $\ell$, we use code
        $W=V_0^\perp$.
        
        [Note: the sets $B$ and $B_1,...,B_n$ are cumulative
        throughout the protocol.]
      \end{nar-enum}
    \item \label{step:l2-frotate} All players apply the inverse
      transform $\F^{-1}$ to their shares of $S_{0,0}$.
    \end{nar-enum}

  \item The remaining shares (i.e. the components of the $n$ systems
    $S_{0,0,i}$) form the sharing of the state $\rho$.
  \end{nar-item}
\end{protocol}
}}}\caption{\protref{vqss} (\vqss---Sharing Phase)}
\label{fig:vqss}
\end{center}
\end{figure}

\clearpage

\begin{figure}[t!]
  \begin{center}
    \fbox{\parbox[t]{\textwidth}{\small
        \begin{protocol}{\vqss---Reconstruction Phase} \label{prot:qr} 
          Player $j$ sends his share of each of the systems
          $S_{0,0,i}$ to the receiver $R$, who runs the following
          decoding algorithm:
          \begin{nar-enum}
          \item For each branch $i$: determine if there is a set
            $\tilde B_i$ such that $B_i \subseteq \tilde B_i$,
            $\abs{\tilde B_i} \leq t$ and the shares of $S_{0,0,i}$
            lie in ${\cal C}_{\tilde
              B_i}$.  \\
            If \emph{not}, add $i$ to $B$.\\
            Otherwise, correct errors on $\tilde B_i$ and decode to
            obtain a system $S_i'$.
          \item Apply interpolation to any set of $n-2t$ points 
            not in $B$. Output the result $S'$.
          \end{nar-enum}
        \end{protocol}}}
    \caption{\protref{qr} (\vqss---Reconstruction Phase)}
    \label{fig:qr}
  \end{center}
\end{figure}

Why is this a secure \vqss\ protocol? We want to show that the protocol
is equivalent to the ``ideal model'', where at sharing time the dealer
sends his secret system $S$ to a trusted outside party, and at reveal
time the trusted party sends $S$ to the designated receiver. To do
that, we will use two main technical claims:
\begin{itemize}
\item Soundness: At the end of the protocol, if the dealer passes all tests then
  there is a unique state which will be recovered by the receiver,
  regardless of any changes made by the cheating players.
\item Completeness (simplistic version): If the dealer is honest, then
  he will pass all tests and the state recovered by the receiver will
  be exactly the dealer's input system $S$.
\end{itemize}

At first, it may not be clear that the claim above for completeness is
really sufficient, since it does not explicitly rule out the adversary
learning any information about the secret system $S$. In fact, at some
intuitive level it \emph{is} sufficient, since any information the
adversary was able to learn would cause a disruption of $S$ (in
general). Nonetheless, a formal proof of security requires a more
sophisticated argument.  We give the more formal proof, based on
simulation, in \secref{l2-sim}.

\subsection{(Informal) Soundness}
\label{sec:l2-sound}

\begin{lemma}\label{lem:l2-sound}
  The system has high fidelity to the following statement: ``Either
  the dealer is caught or measuring all shares in the computational
  (resp.  Fourier) basis would yield a $\tgood$ tree with respect to
  the code $V$ (resp. $W$).''
\end{lemma}

\begin{proof}
  \an{This could be clearer...}  The proof of this lemma follows the
  ideas outlined in the proof of soundness for the subspace projection
  protocol. First, a quantum-to-classical reduction allows us to use
  the soundness of the modified classical protocol from
  \secref{classic}: this gives us that at the end of
  \stepref{l2-vcomp}, either $D$ would get caught or all the systems
  $S_{\ell,0}$ would yield $\tgood_V$ trees if measured in the
  computational basis. After applying the Fourier transformations in
  \stepref{l2-rotate}, all the systems will be $\tgood_V$ in the
  Fourier basis. Subsequent application of linear gates will not
  change that, since they correspond to linear gates in the Fourier
  basis. Finally, applying a second quantum-to-classical reduction
  shows that at the end of \stepref{l2-vfour}, the system $S_{0,0}$
  will be $\tgood_W$ in the computational basis.  Since it is also
  $\tgood_V$ in the Fourier basis, the final rotation in
  \stepref{l2-frotate} will leave it $\tgood_V$ in the computational
  basis and $\tgood_W$ in the Fourier basis.
\end{proof}

Let ${\cal E}$ denote the operator used to encode a state using ${\cal
  C}$. Let $J$ be a set of indices $j$ such that the error operators
$\set{E_j}_{j \in J}$ run over all the syndromes of the code ${\cal
  C}$ (i.e. $J$ contains one representative from each equivalence
class of error operators, and the spaces $\set{E_j\mathcal{C}}_{j \in
  J}$ are orthogonal and span $\ce^{p^n}$). Note that $|J|=p^{n-1}$
since the code is 1-dimensional.

\begin{fact}\label{fact:2basis}
  The following set is an orthonormal basis of $p^{n^2}$-dimensional
  Hilbert space $\ce^{p^{n^2}}$(where $p$ is the size of $F$):
  $$\set{E_{j_1}^{(1)}\cdots E_{j_n}^{(n)} {\cal E}^{\otimes n}
    E_{j_0} {\cal E} \ket{a} \quad : \quad j_0,...,j_n \in J, a \in F
    }$$
  where the superscript $~^{(i)}$ on $E_{j_i}$ indicates that it
  acts on the $\ith$ block of $n$ qupits.
\end{fact}
\begin{proof}\an{Maybe I can just refer to concatenated coding literature.}
  First, notice that these vectors are indeed pairwise orthogonal: for
  a pair of vectors, if any of the indices $j_i\in J$ differ for
  $i\geq 1$, we can distinguish the two states by measuring the
  syndrome of the $\ith$ block of qubits. If none of the $j_i$ differ
  but the indices $j_0$ differ, then we can distinguish the two states
  by correcting all the errors $E_{j_i}^{(i)}$, decoding the resulting
  blocks and measuring the syndrome of the final codeword. Finally, if
  the two states differ only by the choice of $a \in F$, we can
  distinguish them by correcting all errors, decoding and measuring
  the resulting qupit in the computational basis.

  On the other hand, there are $p^{n-1}$ choices for each of the $n+1$
  indices $j_0,...,j_n \in J$ and $p$ choices for $a\in F$. Thus the
  total number of states is $\paren{p^{n-1}}^{n+1}p = p^{n^2}$, and so
  the states must span all of $\ce^{p^{n^2}}$.
\end{proof}

\begin{prop}[Characterizing $\tgood$ trees]\label{prop:char2good}
  The space of trees of qupits which are $\tgood_V$ in the
  computational basis and $\tgood_W$ in the Fourier basis is spanned
  by the states\\ $E_{j_1}^{(1)}\cdots E_{j_n}^{(n)} {\cal E}^{\otimes n}
  E_{j_0} {\cal E} \ket{a}$ where
  \begin{itemize}
  \item $E_{j_0}$ (or something in its equivalence class) acts only on
    $B$ and
  \item For each $i \not \in B$, $E_{j_i}$ (or something in its
    equivalence class) acts only on $B_i\cup C$. \\
    (Recall that for $i$ corresponding to honest players not in $B$,
    we have $B_i \subseteq C$ and so in those cases the condition is
    that $E_{j_i}$ act only on $C$.)
  \end{itemize}
\end{prop}
\begin{proof}
  Given any state of $n^2$ qupits, we can write it as a mixture of
  linear combinations of basis vectors from the basis in the previous
  discussion (\factref{2basis}). Now for any one of these basis states given
  by $j_0,...,j_n$ and $a$, it will pass a test of $\tgood$-ness in
  both bases if and only if the conditions of the proposition are
  satisfied: $j_0$ should be the syndrome of some error which acts
  only on $B$ and each $j_i$ should be equivalent to an error on $B_i
  \cap C$. Thus, any state which passes the test with probability 1
  can in fact be written only in terms of those basis vectors which
  pass the test.
\end{proof}

Note that in the case of the basis vectors of the previous
proposition, there is no entanglement between the data and the errors,
since the data is a pure state (in fact, we can also think of the
errors as being described by a pure state $\ket{j_0,...,j_n}$).
However, one can get arbitrary superpositions of these basis vectors
and so in general there will be not only correlation, but indeed
entanglement between the data and the errors.

\paragraph{Ideal Reconstruction}

In order to prove soundness carefully, we define an \emph{ideal
  interpolation} circuit ${\cal R}^I$ for $\tgood$ trees: pick the
first $n-2t$ honest players not in $B$, say $i_1,...,i_{n-2t}$. For
each $i_j$, pick $n-2t$ honest players not in $B_{i_j}$ and apply the
normal interpolation circuit (i.e.  erasure-recovery circuit) for the
code to their shares to get some qupit $R_{i_j}$. This will yield
$n-2t$ qupits total. Applying the interpolation circuit again, we
extract some system $S$ which we take to be the output of the ideal
interpolation. For simplicity, we assume that the interpolation
circuit extracts the encoded state and replaces it with an encoding of
$\ket{0}$, i.e. it maps ${\cal E}\ket{a} \longmapsto
\ket{a}\otimes{\cal E}\ket{0}$.

\begin{lemma}\label{lem:l2-well-defined}
  Given a tree of qupits which is $\tgood$ in both bases, the output
  of the ideal interpolation and the real recovery operators are the
  same. In particular, this means that no changes made by cheaters to
  their shares of a $\tgood$ tree can affect the outcome of the
  recovery operation.
\end{lemma}

Note that this is not necessarily true for a ``one level'' sharing
(\secref{dualSP}), unless $t<n/8$: by entangling errors with the
shared data, the cheaters could arrange things so that more than $t$
errors are detected only for certain possible values of the data,
creating an entanglement between the data and the success or failure
of the recovery.

\begin{proofof}{\lemref{l2-well-defined}}
  Both the decoding and recovery operators produce an output qubit as
  well as an ancilla. We show that there is a unitary map which can be
  applied to the ancilla of the interpolation operator so that the
  joint state of the output and the ancilla are the same as when the
  decoding operator is applied.

  It is sufficient to prove this for some basis of the space of
  $\tgood$ trees; the rest follows by linearity. The natural basis is
  given by \propref{char2good}. Consider a basis vector
  $E_{j_1}^{(1)}\cdots E_{j_n}^{(n)} {\cal E}^{\otimes n} E_{j_0}
  {\cal E} \ket{a}$ which satisfies the conditions of
  \propref{char2good}.

  \begin{description}
  \item[Effect of ideal recovery] Let $I$ be the set of $n-2t$ indices
    not in either $B$ or $C$, and suppose for simplicity that
    $I=\set{1,...,n-2t}$ (the same argument works regardless of the
    particular values in $I$).  Applying the ideal recovery operator
    to the branches in $I$, we obtain $n-2t$ encodings of $\ket{0}$
    with errors $j_1,...,j_{n-2t}$, and an encoding of $\ket{a}$ whose
    first $n-2t$ positions are untouched and whose last $2t$ positions
    are themselves encoded and possibly arbitrarily corrupted. This
    can be written:
    $$
    \paren{E_{j_{1}} {\cal E}\ket{0}} \otimes \cdots \otimes
    \paren{E_{j_{n-2t}} {\cal E}\ket{0}}\ \otimes\ 
    E_{j_{n-2t+1}}^{(n-2t+1)}\cdots E_{j_n}^{(n)}
    \paren{\mathbb{I}^{\otimes (n-2t)}{\cal E}^{\otimes 2t}} E_{j_0}
    {\cal E} \ket{a}$$
    where $\mathbb{I}$ is the identity. Applying
    ideal recovery again to the first $n-2t$ positions of the encoding
    of $\ket{a}$, we extract $\ket{a}$ and leave a corrupted encoding
    of $\ket{0}$:
    $$\ket{a} \otimes \Big(\paren{E_{j_{1}} {\cal E}\ket{0}} \otimes \cdots
    \otimes \paren{E_{j_{n-2t}} {\cal E}\ket{0}}\ \otimes\
    E_{j_{n-2t+1}}^{(n-2t+1)}\cdots E_{j_n}^{(n)}
    \paren{\mathbb{I}^{\otimes (n-2t)}{\cal E}^{\otimes 2t}} E_{j_0}
    {\cal E} \ket{0}\Big)$$


  \item[Effect of real reconstruction] Now consider the effect of the
    decoding operator, which must be applied without knowledge of the
    positions which are corrupted.  The first operation to be
    performed is to attempt to decode each branch $i \not \in B$. This
    means copying the syndrome $j_i$ for each branch into an ancilla
    state $\ket{j_i}$. Whenever $E_{j_i}$ acts on a set $\tilde B_i$
    such that $|\tilde B_i \cup B_i| \leq t$, then $E_{j_i}$ can be
    identified and corrected.  When $E_{j_i}$ acts on too many
    positions, then it cannot be identified uniquely and the decoding
    procedure will simply leave that branch untouched.

    Let $I$ be the set of indices not in $B$ which had few enough
    errors to correct. At the end of this first phase the input basis
    state will become:
    $$\paren{\big(\prod_{i \not \in I} E_{j_i}^{(i)}\big) {\cal
        E}^{\otimes n} E_{j_0} {\cal E} \ket{a}}\ \otimes\
    \bigotimes_{i \in I}\ket{j_i}$$

    We know that all the honest players not in $B$ are in $I$ (by
    assumption of $\tgood$-ness) and so $I$ contains at least $n-2t$
    positions. Decoding each of these circuits and applying the
    interpolation operator to the resulting qupits, we can extract the
    state $\ket{a}$ and replace it with $\ket{0}$ in the top-level
    sharing. This yields
    $$\ket{a} \otimes \paren{\big(\prod_{i \not \in I}
      E_{j_i}^{(i)}\big) {\cal E}^{\otimes n} E_{j_0} {\cal E}
      \ket{0}}\ \otimes\ \bigotimes_{i \in I}\ket{j_i}$$
  \end{description}

  In both cases, the output can be written as $\ket{a}$ tensored with
  some ancilla whose state depends only on the syndromes
  $j_0,j_1,...,j_n$.  Once that state is traced out, the outputs of
  both operators will be identical. Another way to see this is that
  the ideal operator can simulate the real operator: one can go from
  the output of the ideal operator to that of the real operator by
  applying a transformation only to the ancilla.
\end{proofof}

\lemref{l2-sound} and \lemref{l2-well-defined} together imply that
there is essentially a unique state which will be recovered in the
reconstruction phase when the receiver $R$ is honest. Thus, the
\protref{vqss} is sound, in the informal sense of \secref{vqss}.

\subsection{(Informal) Completeness}
\label{sec:l2-complete}

As discussed earlier, the protocol is considered complete if when the
dealer is honest, the state that is recovered by an honest
reconstructor is exactly the dealer's input state.

\begin{lemma}\label{lem:complete}
  When the dealer $D$ is honest, the effect of the verification phase
  on the shares which never pass through cheaters' hands is the identity.
\end{lemma}

\begin{proof}
  This follows essentially by inspection: for any codeword
  $\mathbf{v}$ of a linear code $W$, applying a controlled addition to
  $\ket{\mathbf{v}}\otimes \sum_{\mathbf{w} \in W}\ket{\mathbf{w}}$
  results in the identity. Since this operation is transversal, the
  shares which never go through cheaters' hands will behave as if the
  identity gate was applied.
\end{proof}

Consider the case where the dealer's input is a pure state
$\ket{\psi}$. On one hand, we can see by inspection that an honest
dealer will always pass the protocol. Moreover, since the shares that
go through honest players' hands only remain unchanged, it must be
that if some state is reconstructed, then that state is indeed
$\ket{\psi}$, since the ideal reconstruction operator uses only those
shares. Finally, we know that since the dealer passed the protocol the
overall tree must be $\tgood$ in both bases, and so some value will be
reconstructed. Thus, on input a pure state $\ket\psi$, an honest
reconstructor will reconstruct $\ket\psi$. We have proved:

\begin{lemma}\label{purestate}
  If $D$ and $R$ are honest, and the dealer's input is a pure state
  $\ket\psi$, then $R$ will reconstruct a state $\rho$ with fidelity
  $1-2^{-\Omega(k)}$ to the state $\ket\psi$.
\end{lemma}

Not surprisingly, this lemma also guarantees the privacy of the
dealer's input. By a strong form of the no cloning theorem
(\secref{qss})
, any information the cheaters could obtain would cause some
disturbance, at least for a subset of the inputs. Thus, the protocol
is in fact also private.

\subsection{Simulatability}
\label{sec:l2-sim}

The previous two sections prove that the protocol satisfies an
intuitive definition of security, namely that it is complete, sound
and private. In this section, we sketch a proof that the protocol
satisfies a more formal notion: it is equivalent to a simple ideal
model protocol. The equivalence is statistical (\defref{statsecur}),
that is the outputs of the real and ideal protocols may not be
identical, but have very high fidelity to one another.


\paragraph{An Ideal Model Protocol}

Now, it is fairly simple to give an ideal protocol for \vqss: in the
sharing phase, the dealer $D$ sends his system $S$ to $\ttp$. If $D$
does not cooperate or sends an invalid message, $\ttp$ broadcasts ``$D$
is a cheater'' to all players. In the reconstruction phase, $\ttp$ 
sends the system $S$ to the designated receiver $R$. This protocol is
in fact given in \protref{vqss-ideal} (p.~\pageref{prot:vqss-ideal}).

Intuitively, this is the most we could ask from a secret sharing
protocol: that it faithfully simulates a lock box into which the dealer
drops the system he wishes to share.

In order to show equivalence of our protocol to the ideal protocol, we
will give a transformation that takes an adversary $\A_1$ for our
protocol and turns it into an adversary $\A_2$ for the ideal protocol.
To give the transformation, we exhibit a simulator $\Sim$ which acts
as an intermediary between $\A_1$ and the ideal protocol, making
$\A_1$ believe that it is in fact interacting with the real protocol.

\subsubsection{Simulation Outline}

We give a sketch of the simulation procedure in \algref{vqss-sim}
(\figref{vqss-sim}). \an{Insert English description here.}

\begin{figure}[tp]
  \begin{center}
    \fbox{ \parbox[t] { \textwidth }{ \small
      \begin{algm} \label{alg:vqss-sim} Simulation for \vqss\ (\protref{vqss})
        \begin{description}
        \item[Sharing/Verification phase] ~
              \begin{itemize}
              \item If $D$ is a cheater, $\Sim$ must extract some
                system to send to $\ttp$:
                \begin{nar-enum}
                \item Run Sharing and Verification phases of
                  \protref{vqss}, simulating honest players. If $D$ is
                  caught cheating, send ``I am cheating'' from $D$ to
                  $\ttp$.
                \item Choose $n-2t$ honest players not in $B$ and
                  apply ideal interpolation circuit to extract a
                  system $S$.
                \item Send $S$ to $\ttp$.
                \end{nar-enum}
              \item If $D$ is honest, $\Sim$ does not need to send
                anything to $\ttp$, but must still simulate the sharing
                protocol.
                \begin{nar-enum}
                \item Simulate an execution of the Sharing and
                  Verification phases of \protref{vqss}, using $\ket
                  0$ as the input for the simulated dealer $D'$.
                \item Choose $n-2t$ honest players (they will
                  automatically not be in $B$ since they are honest)
                  and apply the ideal interpolation circuit to extract
                  the state $\ket 0$.
                \item The honest $D$ will send a system $S$ to $\ttp$.
                \end{nar-enum}
              \end{itemize}
              \textbf{Note:} Regardless of whether $D$ is honest or
              not, at the end of the sharing phase of the simulation,
              the joint state of the players' shares is a tree that is
              (essentially) $\tgood$ in both bases, and to which the
              ideal interpolation operator has been applied.  Let $I$
              be the set of $n-2t$ honest players (not in $B$ or $C$)
              who were used for interpolation.
            \item[Reconstruction phase] ~
              \begin{itemize}
              \item If $R$ is a cheater, $\Sim$ receives the system
                $S$ from $\ttp$. He runs the interpolation circuit
                backwards on the positions in $I$, with $S$ in the
                position of the secret. He sends the resulting shares
                to $R$.
              \item If $R$ is honest, the cheaters send their
                corrupted shares to $\Sim$. These are discarded by
                $\Sim$.
              \end{itemize}
              In both cases, $\Sim$ outputs the final state of $\A_1$
              as the adversary's final state.
            \end{description}
          \end{algm}
        }}
      \caption{\algref{vqss-sim} (Simulation for
        \vqss)}
      \label{fig:vqss-sim}
\end{center}
\end{figure}

\afterpage{\clearpage}

Why does this simulation work?
\begin{itemize}
\item When $D$ is cheating:
  \begin{itemize}
  \item When $R$ is cheating, the simulation is trivially faithful,
    since there is \emph{no difference} between the simulation and the
    real protocol: $\Sim$ runs the normal sharing protocol, then runs
    the interpolation circuit, sending the result to \ttp. In the
    reconstruction phase, $\Sim$ gets the same state back from \ttp,
    and runs the interpolation circuit backwards. Thus, the two
    executions of the interpolation circuit cancel out.
  \item When $R$ is honest, the faithfulness of the simulation comes
    from \lemref{l2-well-defined}: in the real protocol, $R$ outputs
    the result of the regular decoding operator. In the simulation,
    $R$ gets the output of the ideal interpolation. Since the shared
    state has high fidelity to a $\tgood$ tree (by \lemref{l2-sound}),
    the outputs will be essentially identical in both settings
    (i.e. they will have high fidelity).
  \end{itemize}
\item When $D$ is honest:
  \begin{itemize}
  \item To see that the simulation works when $D$ is honest, we must
    show that two versions of the protocol are equivalent: in the
    first version, $\Sim$ gets $S$ \emph{after} having simulated the
    sharing phase with $\A_1$, and so he ``swaps'' it in by first
    running the ideal interpolation circuit, exchanging the system $S$
    for the shared state $\ket 0$, and then running the interpolation
    circuit backwards.

    In the second version, he gets the system $S$ from $\ttp$
    \emph{before} running the simulated sharing phase, and so he
    simply runs it with $S$ as the input for the simulated dealer
    $D'$.
    
    To see that the two versions are equivalent, view the ``swap'' as
    an atomic operation, i.e. view the application of the
    interpolation, switching out the $\ket{0}$ state and replacing it
    with $S$, and reapplying the interpolation backwards, as a single
    step. Now consider moving the swap backwards through the steps of
    the protocol. Because each of the verification steps acts as the
    identity on the shares of the honest players, we can move the swap
    backwards through all verifications (Note: the verification acts
    as the identity only when the dealer is honest, but that is indeed
    the case here).  Finally, one can see by inspection that sharing a
    $\ket{0}$ and then swapping is the same as sharing the system $S$.
    Thus the two versions of the protocol are equivalent, and so the
    simulation is faithful when $D$ is honest.
  \end{itemize}
\end{itemize}

We have essentially proved:

\begin{theorem}
  \protref{vqss} is a statistically secure implementation of verifiable quantum
  secret sharing (\protref{vqss-ideal}).
\end{theorem}

\subsection{Round and Communication Complexity}
\label{sec:l2-complexity}

In this section we show how to reduce the complexity of the protocol.
For now, we will continue to assume that all public coins are
generated using classical \cvss: all players commit to a random value,
then open all their values and take the sum to be the public coin. We
discuss removing this assumption below.

\paragraph{Reducing the Number of Ancillas}

The first observation is that with these cut-and-choose protocols, it
easy to check many trees at once for $\tgood$-ness, so long as they
were all generated by the same dealer. Suppose that we want to verify
$\ell$ trees of quantum shares for $\tgood$-ness in a certain basis.
The dealer distributes the trees, and then creates $k$ sharings of the
ancilla state $\sum\ket{a}$ (as in the original protocol). In the
original protocol, for each ancilla we chose a random coefficient
$b\in F$ and performed the gate $(x,y)\mapsto (x,y+bx)$. In the new
protocol, we add a random linear combination of all $\ell$ states to
be checked into the ancilla: each challenge consists of $\ell$
coefficients $b_1,...,b_\ell$ chosen publicly at random. We apply the
linear gate $(x_1,...,x_\ell,y) \mapsto (x_1,...,x_\ell,y+ \sum b_j
x_j)$ to the $\ell$ trees and the ancilla. The resulting state is then
measured in the computational basis and all players broadcast their
shares.

To ensure good soundness, we can run this protocol $k$ times in
parallel, i.e. using $k$ different ancillas and $k\cdot \ell$
random coefficients (i.e. $k$ challenges of $\ell$ coefficients).
Essentially the same analysis as in the previous sections shows
that at the end of this protocol (with high fidelity) the dealer
will have been caught or the shared states will \emph{all} be
$\tgood$ in the computational basis.

We can use this observation to improve the efficiency of our
\vqss\ protocol. The dealer shares his secret $S$ and also shares
$2k$ ancillas. He uses the first $k$ ancillas to check both the
target state and the remaining $k$ ancillas for consistency in the
Fourier basis. He then uses the remaining ancillas to check the
target state in the computational basis. The number of ancillas
now scales linearly (instead of quadratically) but the protocol
still requires a quadratic number of public values.

\paragraph{Generation of Public Values}

In the preceding discussion we assumed that public values were
truly random. Such truly random coins can be implemented in our
model using classical \cvss, but in fact they need not be. As
pointed out in \cite{CCD88,RB89}, it is sufficient to have players
take turns generating challenges.

Suppose that each player broadcasts $\frac k n$ random challenges,
and all players apply the challenge and measure and broadcast the
result, as before. Then we are guaranteed that at least
$k'=k\frac{n-t} n$ challenges will be chosen truly at random.
Thus, by increasing $k$ by a factor of $\frac n {n-t}$ we get the
same soundness as before, and avoid expensive \cvss\ protocols.

The final protocol takes three rounds, two of which use the
broadcast channel. Each player sends and receives $kn\log |F|$
qubits. Moreover, the broadcast channel gets used $k$ times to
send challenges of (roughly) $k\log |F|$ bits. It is also used to
broadcast $k$ responses of $n\log |F|$ bits. To have soundness
$\epsilon$, we must have the number of truly random challenges be
$k' \geq \frac{n + O(\log n) +log(1/\epsilon)}{\log|F|}$.

Since $\frac n {n-t}$ is constant, we get quantum communication
complexity $O\paren{\frac{(n + \log \frac 1 \epsilon)^2}{n \log |F|}}$
per player and overall broadcast complexity $O\paren{(n + \log \frac 1
  \epsilon)(n + \frac{n + \log \frac 1 \epsilon}{n \log |F|})}$. This
is optimized when each player broadcasts only a single challenge, i.e.
$\log |F| =\frac{n + \log \frac 1 \epsilon}{n}$. In that case, we get
quantum communication complexity $O(n + \log \frac 1 \epsilon)$ per
player and overall broadcast complexity $O\paren{n(n + \log \frac 1
  \epsilon)}$.

\subsection{Additional Properties of Two-Level Sharing}
\label{sec:l2p}

Level 2 sharings produced by the same dealer (using the protocol
above) have some additional properties, which will be useful for
multi-party computation. First of all, notice that there is no problem
in tracking the sets $B,B_1,...,B_n$ across various invocations of the
protocol for the same dealer. Because set $B_i$ corresponds to the set
of players which player $i$ has accused of cheating, we may take these
sets as cumulative, and simply declare that a player is cheating
whenever the union of all the set $B_i$ (for the same $i$) is greater
than $t$. Similarly for the set $B$. Thus, in the discussion below we
assume that the sets $B,B_1,...,B_n$ are the same for all invocations
with a particular dealer.

\begin{nar-enum}
\item Say the systems $S_{i,j}$, $S'_{i,j}$ form valid two-level
  sharings of states $\rho, \rho'$ respectively (where $S_{ij}$
  corresponds to player $j$'s share of branch $i$).

  Then applying the linear operation $(x,y)\to (x,y+bx)$ to the
  systems $S_{i,j}\otimes S'_{i,j}$ results in
  valid two-level sharings of the states obtained by applying the gate
  to the state $\rho \otimes \rho'$.

  In other words, if we denote the reconstruction procedure by
  $\mathcal{R}$ and the controlled-addition by $c\mbox{-}X^b$, we get that
  $$(c\mbox{-}X^b)\mathcal{R}^{\otimes 2} = \mathcal{R}^{\otimes 2}
  (c\mbox{-}X^b)^{\otimes n^2}$$ (at least when restricted to the subspace of
  valid sharings).

\item Say the systems $S_{i,j}$ form valid two-level sharings of state
  $\rho$ with respect to the codes $V,W$. Then applying $\mathcal{F}$
  to each of the shares results in a valid sharing of the state
  $\mathcal{F}\rho\mathcal{F}^\dagger$ with respect to the codes
  $W,V$.
  
  That is, if $\mathcal{R}_{V,W}$ is the reconstruction procedure
  which uses code $V$ in the computational basis and $W$ in the
  Fourier basis, then when we restrict to the subspace of valid
  sharings we get:
  $$ \mathcal{F} \mathcal{R}_{V,W} = \mathcal{R}_{W,V}
  \mathcal{F}^{\otimes n^2} $$

\item If all players measure their shares in a valid sharing of $\rho$
  and then apply classical reconstruction, then they will obtain the
  same result as if they had sent their shares to an honest
  reconstructor and asked him to broadcast the result of measuring
  $\rho$.


  
\item \label{rem:share0} The dealer can use the protocol to
  additionally prove to all players that the system he is sharing is
  the exactly the state $\ket 0$: the ancillas he uses in this case
  will all be sharings of $\ket 0$ (instead of $\sum \ket a$). The
  verification step is the same as before, except now players verify
  that the reconstructed codeword at the top level interpolates to 0.
  
  Similarly, the dealer can prove that he is sharing a state $\sum_a
  \ket a$ by ensuring that all ancillas used for verification in the
  Fourier basis are in state $\ket 0$, and again asking players to
  verify that the reconstructed codeword at the top level interpolates
  to 0 for the checks in the Fourier basis.
\end{nar-enum}

This last point is worth stressing: by tailoring the protocol, the
dealer can verifiably share states $\ket 0$ and $\sum_a \ket a$. This
will be useful for sharing ancillas in the multi-party computation
protocol.


\afterpage{\clearpage} 
\section{Impossibility of \vqss\ when $t \geq \frac{n}{4}$}
\label{sec:imposs}

\begin{lemma}\label{lem:no4}
  No \vqss\ scheme exists for 4 players which tolerates one
  cheater.
\end{lemma}

Before proving this, we need a result from quantum coding theory, on
the relation between error-correction and erasure-correction:

\begin{fact}[$t$-error correction and $2t$-erasure correction] \label{fact:error-erasure}
  Suppose that a quantum code with $n$ components, and dimension at
  least 2 can correct errors on any $t$ positions. Then in fact $\C$
  can correct \emph{erasures} on any $2t$ positions.
\end{fact}

Note that this holds regardless of the dimensions of the individual
components of the code. It also holds when the code in question is a
``mixed state'' code, i.e. some pure states are nonetheless encoded as
mixed states by the encoding procedure. 

It's an interesting and useful property of quantum information that it
cannot be cloned, i.e. there is no procedure which takes an arbitrary,
unknown pure state $\ket \psi$ and replaces it with two exact copies
$\ket \psi \otimes \ket \psi$ (see \secref{qss}). A corollary of this
is that no quantum code with $n$ components can withstand the erasure
of $\ceil{n/2}$ components. If it could, then one could always
separate the codeword into two halves and reconstruct a copy of the
encoded data with each half, yielding a clone of the encoded data. By
the equivalence of $t$-error-correction and $2t$-erasure-correction,
this means that \emph{there is no quantum code that can correct errors
  on any $\ceil{n/4}$ positions}. This is a special case of the
quantum Singleton bound, also called the Knill-Laflamme bound.

\begin{proofof}{\lemref{no4}}
  Suppose such a scheme exists. Consider a run of the protocol in
  which all players behave perfectly honestly until the end of the
  sharing phase. At that point, their joint state can be thought of as
  a (possibly mixed-state) encoding of the secret that was shared. In
  particular, an honest ``receiver'' Ruth, if she were given access to
  the state of all players, must be able to recover the shared state.
  Moreover, she must be able to do so even if one player suddenly
  decides to start cheating and introduces arbitrary errors into his
  state. Thus, the joint state of all players constitutes a
  four-component {\sc qecc} correcting one error.  However, no such
  code exists, not even a mixed-state one, by the quantum Singleton
  bound.
\end{proofof}

\begin{corollary}
  No \vqss\ scheme exists tolerating an adversary structure that
  contains four sets which cover all players.
\end{corollary}
\begin{proof}
  Suppose there exist four disjoint sets $A,B,C,D$ such that $A \cup B
  \cup C \cup D = P$, and a \vqss\ scheme tolerating any adversary
  that can corrupt any one of those sets. Then we can construct a four
  player protocol tolerating one cheater by having each player
  simulate the players in one of the four sets.
\end{proof}

The optimality of our \vqss\ scheme is also an immediate corollary:

\begin{theorem}
  No \vqss\ scheme for $n$ players exists which tolerates
  all coalitions of $\lceil n/4\rceil$ cheaters.
\end{theorem}

Note that we have only proved the impossibility of \emph{perfect}
\vqss\ protocols. However, both the no cloning theorem and the
equivalence of $t$-error-correction and $2t$-erasure-correction hold
when exact equality is replaced by approximate correctness, and so in
fact even statistical \vqss\ schemes are impossible when $t\geq n/4$.


\afterpage{\clearpage} 
\section{Multi-party Quantum Computation}
\label{sec:mpc}

\an{Add running time analysis.}

In this section we show how to use the \vqss\ protocol of the
previous section to construct a multi-party quantum computing
scheme.

First, we give a modified \vqss\ protocol. At the end of the
protocol, all players hold a single qupit. With high fidelity,
either the dealer will be caught cheating or the shares of all
honest players will be consistent in both the computational and
Fourier bases, i.e. there is no set $B$ of ``apparent cheaters''.

\subsection{Level 3 Sharing Protocol}
\label{sec:l3p}

Until now, we have used protocols for tolerating $t<n/4$ cheaters.
However, we are now interested in tolerating $t<n/6$ cheaters. Thus,
we take $n=6t+1$ for simplicity, and as before we set $\delta=2t$
(thus $\delta'=4t$). We will work with the \css\ code $\C$ given by
$V=V^\delta$ and $W=W^{\delta'}$. Recall that this is the \css\ code for
which we have the simple, nearly-transversal fault-tolerant procedures
of \secref{ftqc}. Our goal is to share a state so that at the end all
shares of honest players lie in ${\cal C}_C = V_C^{(q)} \cap \fn
W_C^{(q)}$.

\begin{figure}[ht] 
  \begin{center}
    \fbox{\parbox[t]{\textwidth}{\small
\begin{protocol}{Top-Level Sharing} \label{prot:l3-vqss}
  Dealer $D$ takes as input a qupit $S$ to share.
\begin{itemize}
\item \textbf{Sharing}
  \begin{nar-enum}
  \item \textbf{(Distribution)} The dealer $D$:
    \begin{nar-enum}
    \item Runs the level 2 \vqss\ protocol on input $S$.
    \item For $i= 1,...,\delta$: \\
      Runs level 2 sharing protocol to share state $\sum_a\ket a$
      (see \remref{share0} in \secref{l2p})
    \item For $i=1,...,n-\delta-1$: \\
      Runs level 2 sharing protocol to share state $\ket 0$
      (see \remref{share0} in \secref{l2p})
    \end{nar-enum}
    Denote the $n$ shared systems by $S_1,...,S_n$ (i.e. $S_1$
    corresponds to $S$, $S_2,...,S_{\delta+1}$ correspond to
    $\sum_a\ket{a}$ and $S_{\delta+2},...,S_n$ correspond to $\ket{0}$).
    Note that each $S_i$ is a two-level tree, and thus corresponds to
    $n$ components in the hands of each player.
    
  \item \label{step:l3-comput} \textbf{(Computation)} Collectively,
    the players apply the Vandermonde matrix to their shares of 
    $S_1,...,S_n$. \\
    (If $D$ is honest then system $S_i$ now encodes the $i$-th
    component of an encoding of the input system $S$).

  \item \label{step:l3roll-back} For each $i$, all players send their
    shares of $S_i$ to player $i$.
  \end{nar-enum}
\item \textbf{Quantum Reconstruction} Input to each player $i$ is the
  share $S_i$ and the identity of the receiver $R$.
  \begin{nar-enum}
  \item Each player $i$ sends his share $S_i$ to $R$.
  \item $R$ outputs ${\cal D}(S_1,...,S_n)$ and discards any ancillas.
  \end{nar-enum}
\end{itemize}
\end{protocol}
}}\caption{\protref{l3-vqss} (Top-Level Sharing)}
\label{fig:l3-vqss}
\end{center}
\end{figure}

The new scheme is given in \protref{l3-vqss} (\figref{l3-vqss}).  The
idea is that the previous \vqss\ scheme allows distributed computation
of linear gates and Fourier transforms on states shared by the same
dealer. It also allows verifying that a given shared state is either
$\ket 0$ or $\sum \ket a$. The players will use this to perform a
distributed computation of the encoding gate for the code ${\cal C}$.
Thus, the dealer will share the secret system $S$, as well as $\delta$
states $\sum \ket a$ and $n-\delta-1$ states $\ket 0$. Players then
apply the (linear) encoding gate, and each player gets sent all shares
of his component of the output.

\begin{lemma}
  At the end of \stepref{l3-comput}, the system has high fidelity to
  ``either the dealer is caught or measuring all $n$ trees in the
  computational (resp. Fourier) basis yields a forest of $n$
  $\tgood_V$ (resp. $\tgood_W$) trees whose implicitly defined
  classical values $v_1,...,v_n$ lie in $V$ (resp. $W$).
\end{lemma}
\begin{proof}
   This follows from the linearity of the sharings generated by the \vqss\ scheme.
\end{proof}

\begin{corollary}[Soundness of Top-Level Protocol]
  At the end of the sharing phase (i.e. after
  \stepref{l3roll-back}), the system has high fidelity to ``either
  the dealer is caught or the $n$ shares of players $S_1,...,S_n$ lie
  in ${\cal C}_C$''.
\end{corollary}
\begin{proof}
    This is because the ``rolling back'' of the shares (i.e
    reconstruction of their respective components by all players)
    preserves measurement statistics in both bases.
\end{proof}

\begin{lemma}[Completeness of Top-Level Protocol]
   When $D$ is honest, on input a pure state $\ket{\psi}$, the
   shared state will lie in $\myspan\set{\E\ket{\psi}}_C$, i.e.
   will differ from an encoding of $\psi$ only by a local
   operation on the cheaters' shares.
\end{lemma}

Notice that the dealer can also prove to all players that he has
shared a $\ket 0$ state by simply proving that the system he is
placing in the input position is in state $\ket 0$.

\paragraph{Simulatability and Ideal Secret Sharing}

The top-level protocol (\protref{l3-vqss}) is a simulatable \vqss\ 
protocol, just as was the original protocol. As before, the idea is
that there is no perceivable difference between (\emph{a}) running the
protocol on input $\ket 0$ and having the simulator ``swap in'' the
real shared system $S$ and (\emph{b}) running the protocol honestly.

However, the top-level protocol is also a simulatable implementation
of a different (and stronger) one-phase ideal task, which we call
``ideal secret sharing'' (\figref{iss}). In it, the dealer $D$ sends
his system $S$ to the $\ttp$, and the $\ttp$ encodes it using the quantum
error-correcting code $\mathcal{C}$ and sends the $i$-th component to
player $i$. 

\begin{figure}[ht] 
  \begin{center}
    \fbox{\parbox[t]{\textwidth}{\small
\begin{protocol}{Ideal Secret Sharing}\label{prot:iss}Input: Dealer
  $D$ gets a qupit $S$.
  \begin{nar-enum}
  \item $D$ sends the $|F|$-dimensional system $S$ to $\ttp$. If $D$
    fails to do this, $\ttp$ broadcasts ``$D$ is cheating'' to all
    players.
  \item $\ttp$ encodes $D$ in $\mathcal{C}$. That is:
    \begin{nar-enum}
    \item $\ttp$ creates $\delta$ states $\sum_a \ket a$ and $n-\delta-1$ states
      $\ket 0$.
    \item $\ttp$ runs the linear encoding circuit (given by the
      $n\times n $ Vandermonde matrix) on $S$ and the $n-1$ ancillas.
    \end{nar-enum}
  \item $\ttp$ sends the $\ith$ component of the encoding to Player
    $i$.
  \item For all $i$: Player $i$ outputs either the qupit received from
    $\ttp$ or the message ``$D$ is cheating''. 
  \end{nar-enum}
\end{protocol}
}}\caption{\protref{iss} (Ideal Secret Sharing)}
\label{fig:iss}
\end{center}
\end{figure}

The details of the simulation are substantially similar to those of
\secref{l2-sim}. \an{Should I add anything?}  We get:

\begin{theorem}
  The top-level protocol (\protref{l3-vqss}) is a statistically secure real-world
  implementation of ideal secret sharing (\protref{iss}), for any
  $t<n/4$ (and thus in particular for $t<n/6$).
\end{theorem}

\subsection{Distributed Computation}
\label{sec:dc}

\an{Mention that we do not *need* a composition property, we can run
  the simulation from scratch.}

Given the protocol of the previous section, and given the \ftqc\ 
techniques described in \secref{ftqc}, there is a natural protocol for
multi-party computation of a circuit: have all players distribute their
inputs via the top-level sharing (\protref{l3-vqss}); apply the gates
of $U$ one-by-one, using the (essentially) transversal
implementation of the gates described in \secref{ftqc}; then have
all players send their share of each output to the appropriate
receiver. For completeness, we give this protocol in \figref{mpqc}
(p.~\pageref{fig:mpqc}).

\begin{figure}[hpt] 
  \begin{center}
    \fbox{\parbox[t]{\textwidth}{\small
        \begin{protocol}{Multi-party Quantum Computation}
          \label{prot:mpqc}
          \begin{nar-desc}          
          \item[Pre:] All players agree on a quantum circuit $U$ with
            $n$ inputs and $n$ outputs (for simplicity, assume that the
            $\ith$ input and output correspond to player $i$). The
            circuit they agree on should only use gates from the
            universal set in \secref{ftqc}.
          \item[Input:] Each player gets an input system $S_i$ (of
            known dimension $p$).
          \end{nar-desc}

          \begin{nar-enum}
          \item \textbf{Input Phase:}
            \begin{nar-enum}
            \item For each $i$, player $i$ runs Top-Level Sharing with
              input $S_i$.  
            \item If $i$ is caught cheating, then some player who
              has not been caught cheating yet runs Top-Level Sharing
              (\protref{l3-vqss}), except this time with the
              one-dimensional code $\myspan{\set{\E_\C\ket 0}}$ (i.e. he
              proves that the state he is sharing is $\ket 0$.  If the
              sharing protocol fails, then another player who has not
              been caught cheating runs the protocol. There will be at
              most $t$ iterations since an honest player will always
              succeed.
            \item For each ancilla state $\ket 0$ needed for the
              circuit, some player who has not been caught cheating
              yet runs Top-Level Sharing (\protref{l3-vqss}), with the
              one-dimensional code $\myspan{\set{\E_{\C^\delta}\ket
                  0}}$\ \ or\ \ $\myspan{\set{\E_{\C^{\delta'}}\ket
                  0}}$, as needed. If the protocol fails, another
              player performs the sharing, and so forth.
            \end{nar-enum}
          \item \textbf{Computation Phase:} For each gate $g$ in the
            circuit, players apply the appropriate fault-tolerant
            circuit, as described in \secref{ftqc}. Only the
            measurement used in Degree Reduction is not transversal.
            To measure the ancilla:
            \begin{nar-enum}
            \item Each player measures his component and broadcasts
              the result in the computational basis. 
            \item Let $\mathbf{w}$ be the received word. Players
              decode $\mathbf{w}$ (based on the scaled Reed-Solomon
              code $W^{\delta'}$), and obtain the measurement result
              $b$.
            \end{nar-enum}
          \item \textbf{Output Phase:} For the $\ith$ output wire:
            \begin{nar-enum}
            \item All players send their share of the output wire to
              player $i$.
            \item Player $i$ applies the decoding operator for $\C$
              and outputs the result. If decoding fails (this will
              occur only with exponentially small probability), player
              $i$ outputs $\ket 0$.               
            \end{nar-enum}
          \end{nar-enum}
        \end{protocol}
        }}\caption{\protref{mpqc} (Multi-party Quantum Computation)}
      \label{fig:mpqc}
  \end{center}
\end{figure}

One difficulty in the analysis of this protocol is the measurement
results which are broadcast in the computation phase during Degree
Reduction. If the errors occurring in the measured ancilla were
somehow correlated or entangled with errors in the real data, one
could imagine that measuring and broadcasting them might introduce
further entanglement. However, this will not be a problem: on one
hand, any errors will occur only in the cheaters shares, and so
provide nothing beyond what the cheaters could learn themselves; on
the other hand, the honest players will discard all the information
from the broadcast except the decoded measurement result (each honest
player performs the decoding locally based on the broadcast values, so
all honest players obtain the same result). Again, the cheaters can do
this themselves. A full proof of security is somewhat tedious;
instead, we sketch the main ideas in the remainder of this section.

\begin{lemma}
  Suppose that all inputs and ancillas are shared at the beginning via
  states in $\C_C$. Then the result of applying the protocol for a
  given circuit $U$, and then sending all states to an honest decoder
  $R$ is the same as sending all states to $R$ and having $R$ apply
  $U$ to the reconstructed states.
\end{lemma}

\begin{proof}
  Any state in $\C_C$ can be written as a mixture of linear
  combinations of basis states $E_j \E \ket \psi $ (see
  \lemref{charact}). The works on fault-tolerant computing show that
  the above procedures work correctly on such basis states. More
  importantly, they produce no new entanglement: the only opportunity
  to do so would come from the interaction in the measurement step of
  Degree Reduction. However, the resulting leftover ancilla is
  independent of the data in the computation, and hence provides no
  new information or entanglement. 
\end{proof}

\begin{theorem}  
  For any circuit $U$, \protref{mpqc} is a statistically secure
  real-world implementation of multi-party quantum computation
  (\protref{mpqc-ideal}) as long as $t<n/6$.
\end{theorem}

\begin{proof}
  The proof of this is by simulation, as before. The key observation
  is that when the simulator $\Sim$ is controlling the honest players,
  the adversary cannot tell the difference between the regular
  protocol and the following ideal-model simulation:
  \begin{nar-enum}
  \item $\Sim$ runs the input phase as in the protocol, using $\ket 0$
    as the inputs for honest players. In this phase, if any dealer is
    caught cheating, $\Sim$ sends ``I am cheating'' to the $\ttp$ on
    behalf of that player.
    
  \item $\Sim$ ``swaps'' the cheaters' inputs with bogus data $\ket 0$,
    and sends the data to the $\ttp$. That is, he applies the
    interpolation circuit to honest players' shares to get the various
    input systems $S_i$ (for $i \in \C$), and then runs the
    interpolation circuit backwards, with the state $\ket 0$ replacing
    the original data.
    
  \item $\Sim$ now runs the computation protocol with the adversary on
    the bogus data. (Because no information is revealed on the data,
    the adversary cannot tell this from the real protocol.)
  \item $\Sim$ receives the true computation results destined to
    cheating players from $\ttp$.
  \item $\Sim$ ``swaps'' these back into the appropriate sharings, and
    sends all shares of the $\ith$ wire to player $i$ (again, he does
    this only for $i \in \C$).
  \end{nar-enum}
  The proof that this simulation succeeds follows straightforwardly
  from the security of the top-level sharing protocol and the previous
  discussion on fault-tolerant procedures.
\end{proof}

\afterpage{\clearpage} 
\chapter{Open Questions}
\label{cha:conc}

We conclude briefly with some open questions based on this research:
\begin{itemize}
\item Perhaps the most obvious question, given the results of this
  thesis, is to determine the true threshold for multi-party quantum
  computing, i.e. is it possible to tolerate up to $\floor{(n-1)/4}$
  cheaters? We conjecture that it can indeed be done, but the
  techniques we use here are clearly not sufficient. 
  
  One approach to this problem is to find a fault-tolerant Toffoli
  procedure for the code $\C^\delta$ for $n=2\delta+1$, which
  tolerates $t$ errors at \emph{any} point in the computation. The
  best known procedure for that code is a straightforward
  generalization of Shor's procedure for binary \css\ codes
  \cite{Shor96,AB99}. However, there is one point in that procedure at
  which \emph{at most one} error can be tolerated. Such a procedure
  will fail when $t=\delta/2$ errors can be placed adversarially.

\item A more subtle question is whether or not it is possible to
  remove the error probability from the protocols for verifiable
  quantum secret sharing. Given an error-free implementation of Ideal
  Secret Sharing, error-free multi-party computation is easy. However,
  attaining error-free \vqss\ seems difficult. Although we tried and
  failed to adapt the error-free classical techniques of Ben-Or,
  Goldwasser and Wigderson \cite{BGW88}, we conjecture that it is
  nonetheless possible to achieve error-free quantum computation.
\item A potentially much more difficult question is what tasks are
  achievable when we allow cheating players to force the abortion of
  the protocol. That is, extend the ideal model so that the cheaters
  can, at any time, simply ask the trusted third party to stop the
  protocol entirely. In that setting \vqss\ becomes largely irrelevant
  since an essential aspect of \vqss\ is that the honest players be
  able to reconstruct the secret without the cheaters help. Thus, the
  bound of $n/4$ no longer seems hard; in fact, we conjecture some
  improvement is possible, possibly even up to tolerating any minority
  of cheating players.
\end{itemize}


\appendix

\afterpage{\clearpage} 

\chapter{More on Neighborhoods of Quantum Codes}
\label{cha:morenbhds}

Note that the notions $N_B(\C)$ and $ST_B(\C)$ make sense for any
subspace $\C$ of $\H$. On the one hand, we always have
$N_B(\C)\subseteq ST_B(\C)$ since local operations do not affect the
density matrix of other components. If we restrict our attention to
pure states, then $N_B(\C)$ and $ST_B(\C)$ are in fact identical.
Specifically, define:
\begin{eqnarray*}
  N_B^{pure}(\C) &=& \Big\{ \ket{\psi} \in \mathcal{H}: \exists
  \ket{\phi}\in \C, \exists U\ \mbox{unitary, acting only on
    $\mathcal{H}_B$} 
  \\ 
  && \qquad \qquad \qquad \mbox{such that}\ 
  \ket{\psi} = (I_A\otimes U)\ket{\phi} \Big\} 
  \\ 
  ST_B^{pure}(\C) &=& \Big\{\ket{\psi} \in \mathcal{H}:
    \exists \ket{\phi}\in \C, \tr_B(\ky \bray) = \tr_B(\kf \braf) \Big\}
\end{eqnarray*}

\begin{prop}\label{prop:pure-equal}
  For any subspace $\C$: $N_B^{pure}(\C) = ST_B^{pure}(\C)$
\end{prop}
\begin{proof}
  We must only prove $N_B^{pure}(\C) \supseteq ST_B^{pure}({\cal
  C})$, since the other inclusion is trivial. Take any state $\ky \in
  ST_B^{pure}(\C)$. Let $\kf$ be the corresponding state in
  $\C$ and let $\rho = \tr_B(\ky \bray) = \tr_B(\kf \braf)$. We
  can write $\rho = \sum_i p_i \ket{a_i}\bra{a_i}$ with $p_i>0$,
  $\sum_i p_i =1$, and the vectors $\ket{a_i}$ orthonormal.

  By the Schmidt decomposition, we know that we can write
  $$\ky = \sum_i \sqrt{p_i} \ket{a_i}\otimes \ket{b_i}$$ with the
  vectors $\ket{b_i}$ orthonormal. Similarly, there is some other set
  of orthonormal vectors $\ket{b_i'}$ such that we can write
  $$\kf = \sum_i \sqrt{p_i} \ket{a_i}\otimes \ket{b_i'}$$ Now consider
  any unitary matrix $U$ on $\mathcal{H}_B$ which maps $\ket{b_i'}$ to
  $\ket{b_i}$. Such a matrix always exists since the sets of vectors
  are orthonormal. Then we have $\ky = (I_A \otimes U_B)\kf$ as
  desired.
\end{proof}

This equality does not hold once we relax our definition and consider
mixed states. Namely:

\begin{prop}
  There exist subspaces $\C$ for which $N_B(\C) \varsubsetneq
  ST_B(\C)$.
\end{prop}

\begin{proof}
  Many quantum codes have this property, for an appropriate partition
  of the code word into parts $A$ and $B$.  Take $\C$ to be a quantum
  RS code with $n=2\delta+1$. Encode 1/2 of an EPR pair. Now get
  $\rho$ by appending the other half of the EPR pair to the end of the
  codeword (say there is space left over in position $n$, for
  example). On one hand, $\rho$ is clearly in $ST_B$ as long as $B$
  includes position $n$. However, it is not in $N_B(\C)$ since the
  only state ``in'' $\C$ which has the same trace as $\rho$ on $A$ is
  a mixed state.
\end{proof}

For the case of \css\ codes, we can additionally define $\C_B=\vbq \cap
\fn\wbq$.  Again, there is a trivial inclusion: $ST_B(\C)\subseteq
\C_B$. This inclusion also holds when we restrict our attention to
pure states. However, the inclusion is strict, even for pure states:
\begin{prop}
  There exist subspaces $\C$ for which $ST_B(\C)\varsubsetneq \C_B$.
\end{prop}

\begin{proof}
  Again, consider the quantum RS code with $n=2\delta+1$. Take
  $A=\set{1,...,\delta+1}$ and $B=\set{n-\delta+1,...,n}$. Both $V_B$
  and $W_B$ cover the entire space $\ze_p^n$, so in fact $\C_B$ is the
  entire Hilbert space. However, any state $\rho$ in $ST_B(\C)$ must
  yield $\rho' \otimes I_{\set{2,...,\delta+1}}$ when the
  interpolation operator is applied to the positions of $\rho$ in
  $A$. Thus, not all states, pure or mixed, are in $ST_B(\C)$.
\end{proof}

It should be noted that neither $N_B(\C)$ nor $ST_B(\C)$ are
subspaces. Moreover, for \css\ codes, $\C_B$ is the subspace generated by
the vectors in $N_B(\C)$.

\paragraph{Correspondence to an Idealized Experiment}

One interesting property of $ST_B(\C)$ is that it is exactly the set
of states which will arise in an idealized experiment in which
cheaters introduce errors which are entangled with the data.
Specifically, allow the cheaters to choose an arbitrary joint state
$\ket\psi$ for two systems $L$ and $Aux$ ($L$ is the logical data,
$Aux$ is auxiliary workspace).  Now encode $L$ using $\C$, and allow
the cheaters to apply any operator which affects only $Aux$ and the
components of the encoding contained in $B$. Finally, trace out $Aux$
so that only the components of the (corrupted) codeword are left. 

\an{Insert picture, with the position ``corrupted codeword'' clearly
  identified.}

\begin{prop}
  The set of possible states of the corrupted codeword system in the
  previous experiment is exactly $ST_B(\C)$.
\end{prop}

\begin{proof}
  We can assume w.l.o.g. that the adversary provides a pure state as
  input, since we can always purify the state with an ancilla and have
  him simply ignore the ancilla.  Now, we are in the situation of
  considering $N^{pure}_{B\cup Aux}(\C')$, where $\C'$ is the code
  consisting of $\C$ when restricted to the codeword positions (and no
  restrictions on the $Aux$). By \propref{pure-equal} this is equal to
  $ST^{pure}_{B\cup Aux}(\C')$. But we have $ST_B(\C)=ST_{B \cup
    Aux}^{pure}(\C')$, i.e. once we trace out everything but $A$,
  there is no difference between $\C$ and $\C'$.
\end{proof}

\bibliographystyle{alpha}
\bibliography{crypto}

\end{document}